\documentstyle[12pt,aasms4,epsf]{article}

\begin{document}

\def\date#1{\line{\hbox to 4.0truein {\null\hfill} {#1} \hfill}}
\def\vlsr{V$_{lsr}$}
\def\b-v{$(B-V)$\kern- .15em$_{\circ}$}	
\def\md{\raise 1.7ex \hbox{$\scriptstyle m$}
    \kern- 1.1 .}	
\def\o.{{\rlap.}{^\circ}}    
\def\mdd{\raise 1.7ex \hbox{$\scriptstyle m$} \kern- 1.1em .$\,$}
\def\s.{{\rlap.}{\char'175}}	
\def\m.{{\rlap.}{\char'023}}	
\def\gta{\lower 0.5ex\hbox{$\buildrel > \over \sim\ $}}	
\def\lta{\lower 0.5ex\hbox{$\buildrel < \over \sim\ $}}	
\def\kms{km$\,$s$^{-1}$}
\def\wcen{$\omega$ Cen}
\def\newpage{\vfill \eject}
\def\1sk{\vskip\the\baselineskip}	
\def\oneskip{\vskip\the\baselineskip}	
\def\halp{H$\alpha$}
\def\msun{$M_\odot$}
\def\sec{$^{\prime\prime}$}

\def\refs{\centerline{{\bf References}}\hoffset=-0.25truein}
\def\ref#1\par{\hangindent=6em \hangafter=1
    {#1} \par}	

\def\aap{{A\&A,}\ }
\def\aaps{{A\&AS,}\ }
\def\aj{{AJ,}\ }
\def\apj{{ApJ,}\ }
\def\apjl{{ApJ,}\ }
\def\apjs{{ApJS,}\ }
\def\araa{{ARAA,}
\ }
\def\mnras{{MNRAS,}\ }
\def\pasp{{PASP,}\ }
\def\rmaa{{RMAA,}\ }

\def\strom{Str\"omgren}
\def\filt{$uvby$}
\def\ref#1\par{\hangindent=6em \hangafter=1
    {#1} \par}  
\def\Sec{\hbox{${}^{\prime\prime}$\llap{.}}}
\def\Min{\hbox{${}^{\prime}$\llap{.}}}
\def\ex #1;#2 {$#1\times 10~{#2}\,$}
%
\def\kms{km\ s$^{-1}\,$}
\def\cc{cm$^{-3}\,$}
%
\def\Lsun{L$_\odot\,$}
\def\Msun{M$_\odot\,$}
\def\Rsun{R$_\odot\,$}
\def\mm{$\rm \mu m\ $}
\def\vlsr{V$_{lsr}$}
\def\o.{{\rlap.}{^\circ}}    
\def\s.{{\rlap.}{\char'175}}    
\def\m.{{\rlap.}{\char'023}}    
\def\gta{\lower 0.5ex\hbox{$\buildrel > \over \sim\ $}} 
\def\lta{\lower 0.5ex\hbox{$\buildrel < \over \sim\ $}} 
\def\kms{km$\,$s$^{-1}$}
\def\etal{{\it et~al.}}

\title{\LARGE Age and Metallicity Effects in $\omega$ Centauri:}
\title{\Large Stromgren Photometry at the Main-Sequence Turn-Off}
\author{\Large  Joanne Hughes  \footnote{Visiting astronomer,
Cerro Tololo Inter-american Observatory, operated by the National Optical Astronomy
Observatory under contract to the National Science Foundation.}
\footnote{Now at: Physics Department, Everett Community College, 801
Wetmore Avenue, Everett, WA 98201}
and
 George Wallerstein }
\affil{e-mail contact: hughes@astro.washington.edu}
\affil{Department of Astronomy, University of Washington, Box 351580, Seattle, WA 98195-1580} 

\begin{abstract}
We have observed (with $vby$ filters) a field north of the core of the most massive globular cluster in our galaxy,
$\omega$ Centauri. We have found a correlation of age and
metallicity in a region which avoids the
dense core and the inhomogeneous foreground dust emission shown by the
IRAS satellite. Our observations show that the comparatively metal-rich stars (as
defined by the $(b-y)$ and $m_1$ colors)
are younger than the metal-poor stars by at least 3 Gyr.
This correlation of metallicity
with age suggests that \wcen\ has enriched itself over a timescale of about
3 Gyr,
and possibly longer.
It is remarkable that
ejecta from stellar winds combined with supernovae of type II failed to disperse
the cluster's interstellar matter at an earlier epoch, but were captured by the
cluster instead.
Star formation would have ceased as type
Ia supernovae dispersed the remaining interstellar matter. This work
and other recent evidence suggests that \wcen\ could have been part of a small satellite galaxy in which all the activity occurred before it was
captured by the Milky Way.

Key words: (Galaxy:) globular clusters: individual: 
(\wcen , NGC 6397) --- techniques: photometric

\end{abstract}

\section{Introduction}

The most massive ($3\times 10^6$\Msun ), and flattened ($e=0.8$) 
globular cluster in our galaxy,
$\omega$ Centauri (NGC~5139), is known to show a wide range of metallicity
among both its RR Lyrae stars (Butler, Dickens and Epps 1978) and its red
giants (Norris and Da Costa 1995). The full range of metallicity is
approximately $-2.0<[Fe/H]<-0.5$, with relatively few stars showing
$[Fe/H]>-1.0$. The question of whether this unusual inhomogeneity in
chemical composition is caused by self-enrichment, a chemically-diverse
parent cloud, the capture of \wcen\ from a satellite system, 
or the formation of the cluster by a merger event, has been
the topic of much discussion (see discussion in Vanture, Wallerstein and 
Brown 1994, Norris \etal\ 1997). Additionally, the absolute ages of globular clusters 
are important because 
they provide a lower limit to the age of the universe (Gratton \etal\ 1997).
In the case of \wcen , evidence of the spread in age and/or metallicity 
gives clues to the chemical enrichment of our galaxy.
The origin of the chemical composition diversity in \wcen\ is not only an
intriguing problem itself: its solution may be helpful in developing an
understanding of the evolution of galaxies, especially ellipticals and the
bulges of spirals (which may also show a range in metallicity).

The distribution of stars as a function of metallicity in \wcen\ has been investigated by 
two groups (Norris \etal\ 1996, and Suntzeff and Kraft 1996). The former group
found the peak in the metallicity distribution at [Ca/Fe]=-1.7, and a second
minor peak at [Ca/Fe]=-0.9. The latter group found that a single distribution
fit the data
with a peak at [Fe/H]=-1.7 (which corresponds to [Ca/Fe]=-1.4), and an
asymmetric tail to higher metallicities, but no secondary maximum.
Recently, Norris \etal\ (1997) cited evidence from 400 red giants, correlating
kinematics with metal abundance. However, their data analysis could not differentiate
between the $\omega$ Centauri metal-poor component being older or younger than
the metal-rich component, but the metal-rich and metal-poor stars do 
appear to have substantially different kinematics, suggesting a merger event.
Norris \etal\ (1997) use the Carina dwarf spheroidal system as their example of
a system with about $10^7  M_\odot$ of stars with four components of the
stellar population differing in ages by several billion years. Could $\omega$ Centauri
have been part of a larger (dwarf galaxy) system which merged with the Milky Way 
(Freeman 1993, Dinescu \etal\ 1999)?

Morgan and Lake (1989) model $\omega$ Centauri as a clear case of self-enrichment,
and the question immediately arises as to over what
timescale the enrichment took place. 
The Morgan and Lake model has been extended by Brown \etal\ (1991) and
Parmentier \etal\ (1999).
Can we detect a
correlation between abundance and age in $\omega$ Centauri?
Alcaino and Liller (1987) suggest that
the abundance variations are primordial, since they persist for both
evolved and unevolved stars. Mukherjee \etal\ (1992, using a
$3^\prime \times 5^\prime$
field, $20^\prime$W of the center of the cluster) found evidence for a maximum
age spread of
$2\times 10^9$ years and concluded that the metallicity spread at the turn-off
was consistent with little spread in age (but that their constraint was weak). 
Evans (1983) suggested that the metal-rich stars in \wcen\ were born
with their excesses of s-process species, rather than having generated them
internally.

Since it is difficult to recognize an age-abundance correlation among the
giants or RR Lyrae stars alone, we have analyzed the main-sequence turn-off (MSTO), 
where theoretical isochrones separate according to age (at a given metallicity).
At least two colors (or color differences) must be measured for the stars, one that
reveals the age and one that correlates with $[Fe/H]$; the best system
designed for this is \strom\ photometry. 
The \strom\ intermediate-band system (\strom\ 1966) has four primary filters (\filt ) whose passbands were designed to measure the 
temperature, metallicity and gravity of a star (earlier than spectral-type G). In this paper, we 
utilize the metallicity index $m_1=(v-b)-(b-y)$. The $y$-filter is
usually converted to the Johnson V, and the $(b-y)$-index is related to 
temperature.
The \em uvby \rm filters usually necessitate large numbers of long
exposures on large telescopes (see method detailed in Folgheraiter \etal\ 1995),
mostly because the stars of interest are faint at the {\it u}-band. However,
our method circumvents the need for long {\it u}-exposures\footnote{Folgheraiter
\etal\ (1995) note that the {\it u}-band is at a wavelength where 
little light is emitted by red stars in globular clusters and
that CCDs are not very sensitive here.} with a 0.9-m telescope.
We can reach almost the same
faint-limit as the Mukherjee \etal\ (1992) study with the CTIO 4.0-m
by avoiding the {\it u}-band.
The $m_1$-indices of stars near the
MSTO of a globular cluster are sensitive to metallicity. In the range 
$-2.0<[Fe/H]<-0.5$, the slope of the $m_1$ versus $[Fe/H]$ is $\Delta [Fe/H]
= 0.056m_1$ (Schuster and Nissen 1989). Hence, we expect a total range in $m_1$ of 0.084 magnitudes.

The models of VandenBerg and Bell (1985) and Vandenberg \etal\ (1999) show that, at the MSTO (which is at
$V \sim 18.3 \; mag.$ for $\omega$ Centauri),
the color-magnitude diagram is nearly vertical. At the MSTO the color index
$(b-y)$ is sensitive to age with a slope given by $\Delta (b-y) = 0.010 \Delta
age(Gyrs)$ between 10 and 15 Gyrs. 
Hence, we can determine the metallicity of the stars using 
the $m_1$-index, divide them into groups of relatively low, medium and high
metallicity, and compare them with the appropriate isochrones.
The new VandenBerg \etal\ (1999) isochrones, with increased
$[\alpha/Fe]$ ratio, corrects differences in color which were seen with the
previous models because an enhanced
$\alpha$-element
abundance makes colors redder and magnitudes fainter (Reid 1997).

We have been careful in
selecting our field ($\sim 25^\prime$N of the cluster center)
to avoid the foreground dust emission shown in IRAS skyflux images
(Wood and Bates 1993, 1994), which could be responsible for variable extinction
across the face of the cluster. A y-image (1500s integration) of our \wcen\ field
is shown in Figure~1.

We took exposures in a field away from the cluster (about $1^\circ$ west
of our cluster field, at approximately the same galactic latitude) to 
subtract field stars statistically. The color-magnitude diagram of stars in
the off-cluster field does not show evidence of the cluster population (Figure~3b).
In an effort to determine the effects of observational uncertainties, we utilized 
observations of NGC~6397 (taken during the same 0.9-m run),
which were deliberately  of poorer quality than that for \wcen . 
Hence, we performed tests to see if a single-metallicity cluster might
show an age spread independent of the metallicity spread 
(Anthony-Twarog, Twarog and Suntzeff 1992) and mimic \wcen .

\section{Observations}

The observations were carried out with
the CTIO 0.9-m telescope in May 1996. We used the
Tek~2048~$\#3$ chip (and Arcon 3.3) which gives a pixel size of
0\Sec396 (focal ratio: f/13.5; pixel size: 24$\mu m$). The read
noise was 3.6$e^-$ and the gain in $e^-/ADU$ was 1.6. The instrument-control 
gain was on the $\#4$ setting (which gave full-well at
approximately 65,000 ADU), because our main project involved observations
of globular cluster stars above and below the MSTO and a 
large dynamic range of magnitudes was needed.
The theoretical field size is 13\Min5 $\times $ 13\Min5, but the
\strom\ filters used at the time were $2\times 2$ inches (instead of $3\times 3$
inches for the broadband filters) and caused vignetting, so the effective field of view was
reduced to about 12\Min0 $\times$ 12\Min0. The frames used for the
analysis are listed in Table~1.
This region includes the field used for the Eggen
photo-electric sequence (Plate I, \it Astronomer Royal, \rm ROA  $\# 2$, 1966).

We used \strom\ standards from the E-regions (Kilkenny and Laing 1992), supplemented by some
metal-deficient giants from Anthony-Twarog and Twarog (1994). The images
were flatfielded using twilight sky-flats, accompanied by a sequence of 25 zeros. The 
images were then processed using the IRAF routine QUADPROC (to take into account the
4 amplifiers used on this system).

The FWHM of stars in all the images was 1\Sec4-2\Sec3. 
We used the DAOPHOT program in IRAF (Stetson, Davis and Crabtree 1990) to perform the 
crowded field photometry including two iterations of (DAOFIND-PHOT-ALLSTAR) with
a $4\sigma$-detection threshold on the first pass through the program, then a $5\sigma$-detection limit on the second pass. The thresholds were set to get the 
best possible photometry at the MSTO, not to detect every possible 
stellar image on the frames. 
Point spread functions (PSFs) were constructed using 20-50 uncrowded stars
in each image and the PSF was allowed to vary quadratically across the chip (although the
degree of non-uniformity was small). 
The objects detected by ALLSTAR were then further selected by requiring that 
they have a $\chi^2$ value between 0.5 and 1.5 (so that only \it real \rm stars are picked, avoiding
cosmic rays or extended objects).

Conversion from the instrumental magnitudes was achieved using the standards
mentioned above with the following equations for 05/10/96.

\begin{equation}
(b-y)=0.956(b-y)_i+0.069(b-y)_iX-0.066X-0.069, \; \sigma_{rms}=0.003
\end{equation}
\begin{equation}
V=y_i-4.065-0.059X-0.164(b-y)_iX+0.236(b-y)_i, \; \sigma_{rms}=0.005
\end{equation}
\begin{equation}
(v-b)=0.872(v-b)_i-0.014(v-b)_iX-0.030X-0.323+0.167(b-y)_i, \; \sigma_{rms}=0.005
\end{equation}
Where X is airmass and the subscript $i$ denotes the instrumental magnitudes.

On 05/09/96, 05/12/96 and 05/14/96, we found that for the  \strom\ filters, using the additional
$constant\times (b-y)_iX$ term produced a greater scatter, so the transformations for the
\strom\  observations on these nights  were:

05/09/96:
 
\begin{equation}
(b-y)=1.080(b-y)_i-0.054X-0.121, \; \sigma_{rms}=0.007
\end{equation}
\begin{equation}
V=y_i-3.996-0.123+0.013(b-y)_i, \; \sigma_{rms}=0.007
\end{equation}
\begin{equation}
(v-b)=0.980(v-b)_i-0.104X-0.278, \; \sigma_{rms}=0.010
\end{equation}

05/12/96:

\begin{equation}
(b-y)=1.036(b-y)_i-0.011X-0.136, \; \sigma_{rms}=0.011
\end{equation}
\begin{equation}
V=y_i-3.996-0.146X+0.106(b-y)_i, \; \sigma_{rms}=0.004
\end{equation}
\begin{equation}
(v-b)=0.975(v-b)_i-0.002X-0.395, \; \sigma_{rms}=0.010
\end{equation}

05/14/96:
 
\begin{equation}
(b-y)=1.054(b-y)_i-0.011X-0.100, \; \sigma_{rms}=0.007
\end{equation}
\begin{equation}
V=y_i-3.979-0.121X+0.043(b-y)_i, \; \sigma_{rms}=0.011
\end{equation}
\begin{equation}
(v-b)=0.980(v-b)_i-0.096X-0.255, \; \sigma_{rms}=0.019
\end{equation}

The rms scatter for the transformation equations indicates that the nights
were photometric with the least scatter being on 05/10/96. 
We determined the
aperture correction between the small (5 pixel) aperture used by ALLSTAR and 
the 
larger aperture used on the standards by cleaning and examining the stars
used to make the PSF in each image. 
Before eliminating any stars from the data
set because of bright neighbors or larger errors, we note that after transforming
{\it y} to V, the average uncertainty at 18$^{th}$ magnitude is better than $1\%$, and it is
better than $3\%$ at 21$^{st}$ magnitude.

We used the IMMATCH routines in IRAF to compare lists of objects detected in each frame,
resulting in several data sets in {\it vby}, which we have then analyzed to form a
coherent picture of this region. The \wcen\ working data set was chosen in
the following way. Taking all the \wcen\ frames in Table~1, Figure~2 shows
the uncertainties in $m_1$ (Figure~2a and 2d), $(b-y)$ (Figure~2b and 2e) and $V$ 
(Figure~2c and 2f). The left-hand panels show 10981 sources which had detections in \it any \rm
vby images from Table~1.
The uncertainties are calculated in quadrature from the photon statistics and how well
DAOPHOT did the processing, the aperture corrections and the standard
photometric errors.
The right-hand panels in Figure~2 show the 2147 objects
which were detected in all the frames (except the 300s exposures on 5/9/96), with
the magnitudes and colors calculated as the weighted-means of each individual
detection. We can see that the errors are significantly smaller for this
restricted sample. For the NGC~6397 data, we selected sources (detected
in the frames in Table~1) which had uncertainties, $\sigma_{m_1} \leq 0.05$,
 $\sigma_{b-y} \leq 0.02$ and  $\sigma_{V} \leq 0.015$. Deliberately, we did not use
weighted means here so that we would broaden the width of the upper main-sequence and
MSTO regions.

\section{Cluster Cleaning}

In the absence of kinematic data, we have to rely on a statistical method
to remove non-members from the clusters. To the data selected in Figure~2
for \wcen , we added the shorter exposures (weighted means of detections
in all the 300s frames) in Table~1 to show the giant
branches and as a check
on the cleaning procedure. The original on-cluster data from all 
the exposures in Table~1 is shown in  a color-magnitude diagram (CMD) in
Figure~3a (2473 stars). The CMD for the ``field'' stars is shown in
Figure~3b. Table~2 (given in its entirity in the electronic edition) lists
the input data for the \wcen\ field, Table~3 gives the data for the off-cluster
field, and Table~4 gives the data for NGC~6397. 

To remove non-members from the cluster, we adapted a method used in Mighell,
Sarajedini and French (1998), which also utilized statistical analysis
from Gehrels (1986). We describe the method briefly here, but a detailed
description is given in Appendix~A. For each star in the \wcen\ field, we
counted the number of stars in the CMD which have $(b-y)$ and $m_1$ colors
within max($2\sigma$, 0.1) mag., and V-magnitudes within max($2\sigma$, 0.2) mag. of
the {\it V, (b-y)} and $m_1$ values for the ``cluster'' star.
Then, we took the values of {\it V, (b-y)} and $m_1$ for each ``cluster'' star to the 
list of stars in the off-cluster field and obtained the
number of field stars falling within the same limits, and calculated the
probability that each star in the \wcen\ region was a member (for more details,
see Mighell, Sarajedini and French 1998).
 Then, if a
uniform random number generator gave a probability less than our calculated
probability, the star was accepted as a cluster member. Figure~3c shows the
accepted cluster members as open circles.
The $m_1$ vs. $(b-y)$ color-color plot for the original \wcen\ sample
is shown in Figure~3e, the off-cluster field stars are shown in Figure~3f,
and the clean sample is shown in Figure~3g as open circles. 
The outliers in Figure~3g were examined; we decided
that the cluster population had $-0.2<m_1<0.25$. The final cleaned sample
for \wcen\ (2121 stars) is seen as a CMD in Figure~3d, and a color-color plot in
Figure~3h. Most objects on the giant branches are preserved and the far-outliers
are removed.

We did not observe an off-cluster field for NGC 6397, and even though
this cluster is in a much more field-contaminated region than \wcen\ 
(Anthony-Twarog, Twarog and Suntzeff 1992). We used the same off-cluster field
to remove field stars from NGC 6397; obviously, the off-cluster field is at a 
different galactic latitude, but again we wanted to see if we could discriminate
between a ``contaminated''
single-metallicity cluster and \wcen . Figure~4a shows the original sample from NGC~6397,
Figure~4b shows the off-cluster field and Figure~4c the cleaned data. Examining
the outliers in Figure~4g enabled us to select the final cleaned sample
in the CMD in Figure~4d (747 stars) and  color-color plot in  Figure~4h.
Because of the greater degree of field contamination, this sample will
almost certainly have some non-members remaining although the restriction
that $-0.2<m_1<0.15$ will have removed the majority of them.

Figure 5a shows the histogram of the $m_1$-index for \wcen , which has
a mean of $m_1 = 0.042 \pm 0.066$ (which is the dispersion in the distribution;
considering the uncertainties in the $m_0$ data, intrinsic dispersion in $m_0=0.06$). Figure~5b shows the data from
Mukherjee \etal\ (1992) (unbroken line) and the same data shifted by 0.069
in $m_1$ (dashed line, as they recommended, to fit the standard system). Our histogram
is not significantly different in peak and shape, but we have a larger data set.
The  $m_1$-distribution for NGC~6397 is shown in Figure~5c, which has
mean $m_1 = 0.017 \pm 0.041$ (intrinsic dispersion in $m_0=0.03$). So, the noisy data for the cluster with
lower metallicity does appear more metal poor in $m_1$ and has a distribution width
about $60\%$ that of \wcen\ (which is likely to be an artifact of the larger
uncertainties in the NGC~6397 data set). The off-cluster field $m_1$-distribution is shown in Figure~5d.

\section{Analysis}

We correct for the effect of interstellar reddening on the colors by adopting 
the relation of Crawford and Mandwewala (1976), with error estimated by
Nissen (1981):
\begin{equation} 
E(m_1)=-0.32E(b-y)\pm 0.03,
\end{equation}
hence
\begin{equation}
m_0=m_1+0.32E(b-y)
\end{equation}
and 
\begin{equation}
E(b-y)=0.7E(B-V)
\end{equation}

For $\omega$ Centauri, we adopt $E(B-V)=0.15$, giving $E(b-y)=0.105$, with a distance
modulus of $(m-M)=13.77$ (Dickens and Wooley 1967, which turns out to be close to
the mean of the other studies). Other studies have used
distance moduli ranging from 13.45 to 14.1 (De Marchi 1999, Bates \etal\ 1992, Mukherjee,
Anthony-Twarog and Twarog 1992); we adopt the Dickens and Wooley  distance, and
a mean $E(B-V)$ because
they give the best fit to the most recent isochrones (VandenBerg \etal\ 1999).
The uncertainty in distance modulus of about 0.25 mag. renders an uncertainty
in \it absolute \rm age of 2-3 Gyr with a given set of isochrones (Bergbusch and VandenBerg
1997), but does not affect the \it relative \rm ages derived for stars within \wcen .
For NGC~6397, we adopt $(m-M)=12.13$ and $E(B-V)=0.185$ (Reid and Gizis 1998).
Hence, for \wcen :

\begin{equation}
(b-y)_0=(b-y)-0.105
\end{equation} and
\begin{equation}
m_0=m_1+0.034 
\end{equation}
Henceforth, $m_0$ is the dereddened $m_1$-index.

\section{The $m_1$-Index}

The $m_1$-index is very sensitive to the metallicity in stars
of the color range of the giants, so we can use the metallicity to 
exclude a few stars whose metallicity is clearly outside the distribution.

We now select the objects in our $\omega$ Cen data set around the MSTO. 
Apart from the well-studied relationship between the $m_1$-index
and [Fe/H] for giants (Schuster and Nissen 1989, Grebel and Richtler 1992), we
need a relationship to convert $m_1$ to [Fe/H] for the upper
main-sequence, the MSTO and the sub-giant branch. We use the work from Malyuto (1994), which gives similar results to Schuster and Nissen (1989), but is a more
stable fit to the available data.
Mayluto determined a relationship between $m_1$ and [Fe/H] within the following ranges:

\begin{equation}
0.22 \leq (b-y) \leq 0.38
\end{equation}
\begin{equation}
0.03 \leq m_1 \leq 0.22
\end{equation}
\begin{equation}
-3.5 \leq [Fe/H] \leq 0.2
\end{equation}

Following Malyuto's (1994) notation, we use:
\begin{equation}
(B-Y)=((b-y)-0.22)/0.16 + 1
\end{equation}
\begin{equation}
M_1=(m_1-0.03)/0.19 + 1
\end{equation}
\begin{eqnarray}
[Fe/H]=5.7071(B-Y)M_1 - 49.9162(B-Y)logM_1 + 7.9971(B-Y)^2logM_1 \nonumber \\ -
 0.5895(B-Y)^3  - 24.0889(1/M_1)
+ 14.6747
\end{eqnarray}
Malyuto determined an uncertainty of $\sigma_{[Fe/H]}=0.15$.

Grebel and Richtler's (1992) calibration for giants is:
\begin{equation}
[Fe/H]={{m_1+a_1(b-y)+a_2}\over{a_3(b-y)+a_4}}
\end{equation}

where
$a_1=-1.24\pm 0.006, \; a_2=0.294\pm 0.03, \; a_3=0.472\pm 0.04, \; a_4=-0.118\pm 0.02$, and equations
(18)-(24) imply \it intrinsic \rm (unreddened) colors, which we diagram as
$m_0$. 

 The loci of equal metallicities in the $(b-y)$ vs. $metallicity$ diagram are well approximated by straight lines
within certain color ranges, 
but not $0.26 \leq (b-y) \leq 0.36$ (the MSTO), which interests
us. The straight-line approximation is valid for dwarfs for $0.4 \leq (b-y) \leq 0.8$.
For giants and supergiants, the range is $0.4 \leq (b-y) \leq 1.1$ Grebel and Richtler (1992) show that
there is no difference in the behavior of giants and supergiants.
Figure~6 shows the 2121 stars in the ``clean'' \wcen\ sample, along with the
straight-line fits to the giants from  Grebel and Richtler (1992) and the 
heavy curves from the Malyuto (1994) calibrations for the MSTO stars.

Figure 7 shows a plot of $m_0 \; vs. \; [Fe/H]_{calc}$ from our data, using
Equation (23), for the objects in the MSTO
region of \wcen\ (Figure~7a) and NGC~6397 (Figure~7b). 
We chose the objects at the
MSTO in both clusters fitting the calibration ranges. It appears that about a
quarter of
the \wcen\ stars plotted in Figure~7a show a noticeably asymmetric scatter in $m_0$, away from the
calibration line, indicating that for these stars, something other than $[Fe/H]$
affects the $m_0$-index. In contrast, the NGC~6397 stars in Figure~7b appear well fitted by the
Malyuto calibration. This effect on the metallicity index in \wcen\ has been noted before
by Mukherjee \etal\ (1992) and was mentioned in Schuster and Nissen (1989). The former
authors suggest that it is due to an enhancement of the CN-bands in the stars with
relatively high $m_0$ values. Even 47~Tuc, which shows no evidence of a metallicity spread,
exhibits a spread in CN-strength extending to the MSTO (Briley, Hesser and Bell 1991).
It seems surprising that moderately metal-poor stars should show an enhancement
of a molecule consisting of two heavy elements since the concentration of such
diatomics should depend on the square of the metallicity. However, it is 
distinctly possible that the atmospheres of stars near the \wcen\ MSTO could be polluted by 
the ejecta of Wolf-Rayet stars, which are greatly enhanced in C and N, but not with
the heavier species. If \wcen\ was able to retain the ejecta from supernovae (Morgan
and Lake 1989), then the ejecta from Wolf-Rayet stars should also be retained. This
effect could explain the 47~Tuc observations, where the cluster was not massive enough
to retain supernovae ejecta.

It is known that
\wcen\ has CN-abundance 
anomalies amongst its giants (Suntzeff and Kraft 1996, Norris and DaCosta 1995), which can affect the metallicity index; the presence of the CN-band
extending to the violet from 4216\AA \    
reduces the flux in the {\it v}-band and pushes the
$m_1$-index higher, which does not mean that the overall metallicity has
changed (Richter \etal\ 1999, Anthony-Twarog \etal\ 1995). As stated above, 47 Tuc, which doesn't
appear to have a dispersion in [Fe/H] on the giant branch,
shows CN-variations at the MSTO, which is thought to be a surface-contamination
 effect
(Cannon \etal\ 1998).

To avoid selecting stars which may not have a simple calibration of $m_0$
to metallicity, we excluded the stars falling away from the calibration line
in Figure~7.
This process also excludes most binaries. The 739 objects 
which met our selection criteria are shown as filled circles in Figure~7a
(rejected objects are shown as open circles). In NGC 6397, 462 stars passed
the selection criteria (Figure~7b). In NGC~6397, which is not known to have
anomalous abundances (Anthony-Twarog, Twarog and Suntzeff 1992), the scatter
is probably due to a larger $\sigma_{m_0}$, binarity and/or overlapping PSFs.

\section{Discussion and Conclusions}
\subsection{The Age-Metallicity Relation}

We want to look for an age spread at the MSTO in addition to \wcen 's
metallicity spread, so we must take care to disentangle the effects of age
and chemical composition. 
By removing
stars from our samples which did not fit the calibration equation, we should have
removed the effect of anomalous abundances and multiple sources (Schuster
\etal\ 1996). 
Figure~8 shows the 10~Gyr isochrones for a range in
metallicities as modeled by VandenBerg \etal\ (1999). \it If stars are all the
same age, the redder objects should have a higher metallicity.  \rm\

First, we compare the \wcen\ stars to the isochrones corresponding
to the mean (spectroscopic) metallicity, $[Fe/H]=-1.54$ in Figure~9. These isochrones are the best fit to the upper
main-sequence and MSTO, but are not a good fit to the subgiants.
This is a consequence of using the Kurucz (1992) colors as a conversion,
tied to the MSTO region (VandenBerg: private communication).
We were encouraged that the latest $\alpha$-enhanced isochrones suggested a lower age for both clusters,
in line with the new \it Hipparcos \rm results (Reid 1997). In Figure~9,
we now divide the selected MSTO sample into
three metallicity groups:
 Figure~9a shows relatively high-metallicity objects: $-1.2<[Fe/H]<-0.5$ (filled triangles), 
Figure~9b shows the medium-metallicity group:
$-1.6<[Fe/H]<-1.2$ (open squares), and
the low-metallicity stars are shown in Figure~9c:  $-2.2<[Fe/H]<-1.6$ (open circles). 
($\sigma_V = \sigma_{b-y} \leq 0.01 $, and $ \sigma_{m_1} \leq 0.015$). 
Clearly, the
relatively metal-rich group (filled triangles) in Figure~9a appear younger than
the metal-poor group (if we assume the whole population is described by 
the $[Fe/H]=-1.54$ isochrones. In this example, the group with
$-1.2<[Fe/H]<-0.5$ has a mean age of $9.7\pm 1.8$~Gyr; the
$-1.6<[Fe/H]<-1.2$ group has a  mean age of $10.2\pm 1.3$~Gyr; and
the $-2.2<[Fe/H]<-1.6$ has a mean age of $11.1\pm 1.2$~Gyr. So, in this
case the age range would be about 1.4~Gyr, with the metal-content increasing
towards younger stars (as one would expect with self-enrichment). 
The NGC 6397 stars are all consistent with being the same age, we find
the mean for the whole NGC~6397
sample is $12.0 \pm 0.7$~Gyr;
any decrease of age with increasing metal-content
is hidden by the uncertainties.

Now, we examine the relatively high-, medium- and low-metallicity groups
in \wcen\ with isochrones of \it different \rm  metallicities, corresponding to
$\sim $ the mean [Fe/H] within each group (Figures~9a--c).
The isochrones in Figure~10a, which contains the most metal-rich stars,
are not a good
fit to the data but indicate an age of about 8~Gyr for the stars with the best
photometry. Figure~10b shows the mid-range group to have a mean age of
$9.4\pm 1.4$~Gyr, and Figure~10c shows the low-metallicity group to have a mean
age of $12.7\pm 1.5$~Gyr. So, Figure~10 implies that there is 
a 2--4~Gyr age range in \wcen .
From Figure~5, there is no strong evidence of a bimodal metallicity
distribution from the $m_1$-index. 
This may be a CN-abundance effect which is gradually
reduced as the stars undergo mixing during later evolution (discussed in
Cannon \etal\ 1998). 

Figure 11a, b and c show the NGC~6397 stars selected in the same way as the 
\wcen\ stars in Figures~9 and 10, with the  isochrones corresponding to a metallicity of
$[Fe/H]=-1.84$. These stars are not well-described by any other set of isochrones, and there
are few stars with good photometry ($\sigma_V = \sigma_{b-y} \leq 0.012$, and $ \sigma_{m_1} \leq 0.023$)
in the apparently high-metallicity group (Figure~11a), as we expect from a cluster with mostly metal-poor
stars. Most of the scatter in Figure~11 is caused by the photometric uncertainties, and the
cluster stars appear to be just under 12~Gyr-old on this set of isochrones. 

In order to explore the relationship between age and metallicity (if any), 
we split the stars in both clusters
into narrow ranges in apparent metallicity: $[Fe/H] = -1.0 \pm
0.1; -1.15 \pm 0.1; -1.3 \pm 0.1; -1.4 \pm 0.1; -1.55 \pm 0.1; -1.6 \pm 0.1;
-1.7 \pm 0.1; -1.82 \pm 0.08; -2.0 \pm 0.1; -2.15 \pm 0.1$. 
We examined the
distribution of mean age with calculated [Fe/H] for \wcen\ (Figure~12a), and
NGC~6397 (Figure~12b). The latter cluster's stars can only be fit well using
the isochrones corresponding to the mean metallicity $[Fe/H]=-1.84$.
The age spread implied by Figure~12a is
at least 3~Gyr and could be as much as 4~Gyr for \wcen . 
NGC~6397 shows no obvious trend of age with metallicity within the uncertainties,
with the stars having a weighted mean age of 12~Gyr (dominated
by points with the smallest error bars). 
NGC~6397 is
more metal-poor than \wcen , as expected, and is described by a 
single-metallicity
and a single age. \wcen\ appears to have a mixture of relatively old metal-poor
stars and a younger metal-rich population, although some of these stars
could have suffered surface contamination. 

\subsection{Possible Histories of \wcen }

We have found that there is an age range of at least 3 Gyr in \wcen ,
and that the stars with higher metallicity are younger; this evidence supports
the self-enrichment theory. Hilker and Richtler (1999) have found that \strom\ 
photometric observations of 1400 red giants show that the age spread between the metal-poor
and metal-rich stars is about 3~Gyrs, with the latter also being younger.
Hence the age spread holds for both the evolved and unevolved stars.

      We now look into the origin of the observed age-metallicity correlation.
Three possibilities were mentioned in the Introduction: 
\begin{enumerate}
\item A chemically diverse
parent cloud; 
\item A merger of two clusters; 
\item Self-enrichment within a single
entity.
\end{enumerate}
      We can easily eliminate the first possibility. While a chemically diverse 
parent cloud -- one in which the dust-to-gas ratio varies within the cloud --
could account for the metallicity spread, there is no reason for such a model to
yield an age-metallicity relation. The possibility that \wcen\ is a merger
has been suggested by Norris \etal\ (1997) on the basis of a second peak in the
metallicity distribution and an apparent difference in the kinematics
of the metal-poor and metal-rich stars. The wide spread in metallicity,
confirmed by Norris and Da Costa (1995) indicates that both of the merging 
clusters must have had an original metallicity spread. In addition the 
correlation of s-process elements with metallicity, reaching $[s/Fe] = 1.0$ at 
$[Fe/H] =
-1.3$ and $[s/Fe] = 1.5$ at $[Fe/H] = -0.7$ shows that the chemical history of 
\wcen\ (or its two progenitor clusters) has been different from that in any other
globular cluster in the Galaxy so far analyzed (Vanture \etal\ 1994). In other
words, both the metal-poor and the metal-rich clusters behaved in a very unusual
 way with regard to their s-process enrichment. These considerations point to
self-enrichment in a single object.
      
In discussing self-enrichment in globular clusters, Morgan and Lake (1989)
require that 330 supernovae of type~II must have provided \wcen\ with 165\Msun\
of iron. The mass limit needed to contain the supernovae  and retain the
ejecta was calculated to be $2.9\times 10^6$\Msun , which is close to the total mass of
the cluster today. Morgan and Lake's calculations suggest that the primordial
cloud from which \wcen\ formed did not have to be 10-100 times as massive
(Padoan, Jiminez and Jones 1997). Having the star formation continue for a free-fall
time ($10^7$~years) all over the cloud to enrich its gas with iron would
reduce the time of star formation well below the level of detectability.
      
To enrich the cluster in s-process elements \wcen\ needed time to
form stars of 2-8\Msun\ which then dumped enriched material into the cluster's
interstellar gas during their AGB phase. This process should have taken about
a billion years and must have been completed before the more powerful Type~Ia 
supernovae,
or a collision with a Galactic gas cloud, swept out the rest of the gas from the
cluster. It seems that the greater mass of \wcen , as compared with other
globulars in our Galaxy was able to retain the enriched gas long enough to
prolong the star formation process. Our data suggest that the oldest, metal-poor, stars are at least 3 Gyrs older than the youngest, metal-rich, stars. In
contrast to other globular clusters, star formation in \wcen\ must have
survived the formation and explosion of the first massive stars (on a time-scale
 of $10^6$ to $10^7$ years) which should have expelled the gas from less massive
globulars.

    For how long did SNIIs continue to contribute metals to \wcen ? The 
abundances of Norris and Da Costa (1995) show that the $\alpha$-elements; Mg,
Si, Ca,
 and Ti, are enhanced with respect to iron all the way from $[Fe/H] = -2.0$ to
$[Fe/H] = -0.7$. Since a high $[\alpha/Fe]$ ratio is associated with SNIIs, 
the supernovae were still contributing all these species while the s-process 
elements were building up from the contributions of 2-8\Msun\ stars. It appears
that SNIas  never contributed much iron to the cluster or else the  value of
$[\alpha/Fe]$ would have diminished as $[Fe/H]$ approached its maximum value. either
the cluster was swept clean by the first SNIas or it lost its interstellar
matter with Galactic gas clouds about when $[Fe/H]$ reached -0.7.
      
In agreement with Suntzeff and Kraft (1996), we find no strong evidence of bimodality in
the distribution of the $m_1$-index, but there is some evidence in the
plot of age vs. metallicity. This evidence alone cannot rule between the
merger of formed clusters or two bursts of star formation;
also, the evidence does not rule out the accretion of another proto-globular
cloud into \wcen\ early in its history (Norris \etal\ 1997). 
The fact that \wcen\ did not lose its interstellar matter though interaction
with Galactic gas clouds for several billion years is difficult to 
imagine if it has always been in its present orbit (Dinescu \etal\ 1999). The
only way that it could have avoided doing so was to be in an orbit far from
the plane of the Milky Way. If this was so, \wcen\ must have been a satellite
galaxy (or the nucleus of a galaxy) that was captured about 8~Gyrs ago. This 
possibility has also been suggested by Dinescu \etal\ and by Majewski \etal\
(1999).

\begin{acknowledgements}

We would like to thank Guillermo Gonzalez, Ken Mighell, John Norris, Peter Stetson, Don VandenBerg, George Lake 
 and Barbara
Anthony-Twarog for useful discussions, and the support staff at CTIO for extra help
in dealing with large-format images.
\end{acknowledgements}

\appendix

\section{``Cleaning'' the Cluster}

This method of ``cleaning'' the cluster membership has been adapted from
Mighell, Sarajedini and French (1998). Since we have separate images of the
on-cluster and off-cluster fields, our use of this method is more straightforward
than their formalism. In addition, instead of having a CMD ($V, \; (b-y)$, Figure~3a)
alone, we
also have a color-color plot ($m_1,  \; (b-y)$, Figure~3e). For each star in the on-cluster
field CMD (Figure~3a), we count the number of stars in the CMD which have
 V-magnitudes within max($2\sigma$, 0.2)~mag., and $(b-y)$ and $m_1$ colors
We call this number
$N_{on}$. Now, we also count the number of \it field \rm stars, in the
off-cluster image, which fall within the same ranges in $V, \; (b-y)$,
(Figure~3b) and $m_1$ (Figure~3f), 
that were determined for the star in the \it cluster \rm field. We call this
number $N_{off}$.  We calculate the probability that the star in the
on-cluster field CMD is a member of the globular cluster population as:
\begin{equation}
p\approx 1-min\left({\alpha N_{off}^{UL \; 84}\over N_{on}^{LL\; 95}},\; 1.0\right)
\end{equation}
where $\alpha$ is the ratio of the area of the
cluster region to the area of the field region, and
\begin{equation}
N_{off}^{UL \; 84}\approx (N_{off}+1)\left[  1- {1\over 9(N_{off}+1)}
+{1.000\over 3 \sqrt{N_{off}+1}}\right ]^3
\end{equation}
(corresponding to eq.(2) of Mighell, Sarajedini and French 1998, and
eq.(9) of Gehrels 1986), is the estimated upper ($\sim 84\%$) confidence
limit of $N_{off}$, using Gaussian statistics.

Now, 
\begin{equation}
N_{on}^{LL\; 95}\approx N_{on}\times 
\left[1-{1\over 9N_{on}} - {1.645\over 3 \sqrt{N_{on}}}
+0.031N_{on}^{-2.50}\right]^3
\end{equation}
is the lower $95\%$ confidence limit for $N_{on}$ (eq.(3) of 
Mighell, Sarajedini and French 1998, and
eq.(14) of Gehrels 1986). 
For \wcen , we assume that
the whole on-cluster field is part of \wcen\ (a fairly safe assumption), so that $\alpha $ is assumed to be
$\sim 1$, in this case.

To estimate if any particular star is a cluster member, we generate a
uniform random number, $0\leq p^\prime \leq 1$, and if $p^\prime \leq p$, we
accept the star as a member of the cluster. The cleaned cluster stars for
\wcen\ are shown as a CMD (2154 objects) in Figure~3c and as a color-color plot in 
Figure~3g. This is a probabilistic method, so the resulting CMD  has to be considered
to be \it ``one of an infinite number of different possible realizations''
\rm (Mighell, Sarajedini and French 1998) of the cleaned cluster population.
Few field stars are found in the MSTO region of the CMD, whereas many 
contaminate the region of the giant branches; therefore the MSTO may be
undercleaned and the evolved stars may be overcleaned. We considered the color-color
plot in Figure~3g, carefully looking at the positions of the outlying stars.
We consider the cluster sequences to lie between $-0.2<m_1<0.25$, rejecting
33 additional objects, but preserving the horizontal branch and possible
blue stragglers.

\newpage
\centerline{\Large Figure Captions}

{\bf Figure 1:}{ A $12.5\times 12.5$ arcminute Str\"omgren y-frame of a field north of the center
of $\omega$ Centauri with a 1500-second integration time. The field center is at $\alpha = 13^h26^m46.0^s$ and $\delta =
-47^\circ 15^{'}00^{\prime\prime}$, which contains the field 
used for Eggen's photoelectric sequence (\it Astronomer Royal, \rm ROA  $\# 2$, 1996)
Number~2, \rm (1966). The frame has been oriented so that north is at the top 
and east is to the left.}

{\bf Figure 2:}
{Left panels: all objects in the $\omega$ Cen field with a detection at any 3 vby frames (10981 sources)
on 05/10/96 (from y: $3\times 900s\; \&\; 1\times 1500s$, b: $3\times 900s\; \&\; 1\times 1800s$,
and v: $3\times 900s\; \&\; 1\times 2000s$, see Table~1): 
\bf a: \rm Uncertainty in the $m_1$-index vs. V magnitude.
\bf b: \rm Uncertainty in the $(b-y)$-index vs. V magnitude.
\bf c: \rm Uncertainty in the $V$-magnitude vs. V magnitude.
Right panels: objects detected in \it all \rm the above frames (2147 objects) on 05/10/96 (see Table~1)
with the V, $(b-y)$ and $(v-b)$ values calculated as the weighted mean of multiple 
detections:
\bf d: \rm Uncertainty in the $m_1$-index vs. V magnitude.
\bf e: \rm Uncertainty in the $(b-y)$-index vs. V magnitude.
\bf f: \rm Uncertainty in the $V$-magnitude vs. V magnitude.}

{\bf Figure 3:}
\bf a: \rm 2473 objects in \wcen\ (2147 from the longer frames in Table~1 plus the 
weighted-mean data from the 300-second exposures, tracing the giant branches).
\bf b:  \rm 536 objects from the off-cluster field.
\bf c: \rm 2154 objects after the first cleaning process shown as open circles.
\bf d: \rm 2121 objects in the final sample after limiting the
$m_1$-index to $-0.2<m_1<0.25$.
\bf e: \rm $m_1 \; vs. \; (b-y)$ color-color diagram for the original 2473
objects in the \wcen\ field. The straight-line fits are for the giants, 
taken from Grebel and Richtler (1992).
\bf f: \rm $m_1 \; vs. \; (b-y)$ color-color diagram for the off-cluster field
containing 536 sources.
\bf g: \rm $m_1 \; vs. \; (b-y)$ color-color diagram for the 2154 stars in \wcen\ 
after the first cleaning pass.
\bf h: \rm $m_1 \; vs. \; (b-y)$ color-color diagram for the 2121 stars in \wcen\ 
after rejecting the far outliers, which preserves the horizontal branch
and possible blue stragglers. Stars with  very low $m_1$ may have colors
affected by an unseen hot companion.

{\bf Figure 4:}
\bf a: \rm 976 objects in NGC 6397 from the $3\times 600s$ exposures at y, and the 
900s exposures at b and v. 
\bf b:  \rm 536 objects from the off-cluster field (same field as used for \wcen\ to test
the cleaning procedure).
\bf c: \rm 808 objects after the first cleaning process shown as open circles.
\bf d: \rm 747 objects in the final sample after considering limits on the
$m_1$-index.
\bf e: \rm $m_1 \; vs. \; (b-y)$ color-color diagram for the original 976
objects in the NGC~6397 field.
The straight-line fits are the loci of equal-metallicity giants from Grebel
and Richtler (1992).
\bf f: \rm $m_1 \; vs. \; (b-y)$ color-color diagram for the off-cluster field
containing 536 sources.
\bf g: \rm $m_1 \; vs. \; (b-y)$ color-color diagram for the 808  sources
after the first cleaning pass.
\bf h: \rm $m_1 \; vs. \; (b-y)$ color-color diagram for the 747 \wcen\ stars
after rejecting the outliers in \bf f\rm . 

{\bf Figure 5:}
\bf a: \rm Histogram of the $m_1$-index for the 2121 objects in the cleaned
\wcen\ sample. The sample mean is $m_1 = 0.042 \pm 0.066$, and we fit a
gaussian to the histogram with those parameters.
\bf b: \rm Histogram of the $m_1$-index for the Mukherjee \etal\ (1992)
sample (solid line) and shifted by their recommended correction (dashed line). 
The histogram of shifted data is not significantly different in shape
from our sample (apart from their truncation in $m_1$), but we have a larger sample of objects. 
\bf c: \rm  Histogram of the $m_1$-index for the 747 objects in NGC 6397. The 
cluster mean is $m_1 = 0.017 \pm 0.041$.
\bf d: \rm Histogram of the $m_1$-index for the 536 objects in the off-cluster
field.

{\bf Figure 6:}
$m_0 \; vs. \; (b-y)_0$ color-color diagram for the 2121 stars in the 
final clean \wcen\ sample. We show the Grebel and Richtler (1992) metallicity
calibration lines (a straight-line approximation) for giants, and the 
Malyuto calbration curves (heavy lines) for the MSTO stars.

{\bf Figure 7:}
\bf a: \rm Plot of reddening-free metallicity index, $m_0$ in \wcen , against the
calculated metallicity from Malyuto (1994) for the ranges: 
$
0.22 \leq (b-y) \leq 0.38
$;
$
0.03 \leq m_1 \leq 0.22
$;
$
-3.5 \leq [Fe/H] \leq 0.2
$.
(We use the Maluto (1994) calbration from Equation~(23), but the Schuster and Nissen (1989)
calibration does not make a great difference to the calculated [Fe/H] values).
There were 1124 sources in \wcen\ within this range and
the filled circles are the objects  (853) which are a good fit to the calibration line.
The open circles are sources which fell in the correct color range, but were
not a good fit. Reasons for some stars not being well-fit by the calibration
 could be anomalous $m_0$-values owing to
enhanced CN-abundance, strong G-bands, measurement errors or binarity/merged images
 (Schuster
\etal\ 1996). The scatter is most pronounced at the high-metallicity end,
supporting the idea some kind of CN-enhancement is probably responsible. 
\bf b: \rm Plot of reddening-free metallicity index, $m_0$ in NGC 6397, 
against the calculated metallicity from Malyuto (1994).  There were
541 within the Malyuto (1994) calibration range and 462 were good fits
to the line. 
As we expect, there is less scatter in this plot because NGC 6397 is not known to have
abundance anomalies. The outliers are most likely caused by binarity and
merged images here, so this
method is a way to remove these objects from the sample; the true binary fraction in
NGC 6397 could be around $10\%$ (Alcaino \etal\ 1997), while it is
considered to be lower in \wcen\ (Mukherjee \etal\ 1992).

{\bf Figure 8:}
Color-magnitude diagram showing the effect of varying the input
metallicity on the color of the main-sequence turn-off. We show the 
10 Gyr isochrones for various metallicities from VandenBerg \etal\ (1999).
If stars are the same age, the redder objects should be more metal-rich.

{\bf Figure 9:}
Color-magnitude diagrams of the main-sequence turn-off region of
\wcen , shown with the isochrones with  $[Fe/H]=-1.54$. We show the stars
divided into groups of relatively high, medium and low metallicity.
All objects shown as filled triangles, open squares and open circles have
$\sigma_V = \sigma_{b-y} \leq 0.01$, and $ \sigma_{m_1} \leq 0.02$.
\bf a: \rm filled triangles have $-1.2<[Fe/H]<-0.5$, and have a mean age of 
$9.7\pm 1.8\;  Gyr$; \bf b: \rm
open squares have $-1.6<[Fe/H]<-1.2$, and have a mean age of 
$10.2\pm 1.3\;  Gyr$; \bf c:
\rm
open circles have $-2.2<[Fe/H]<-1.6$, and have a mean age of 
$11.1\pm 1.2\;  Gyr$. So, if all the stars are well-fit by this set of isochrones,
the stars which are relatively more metal-rich are younger than the more
metal-poor stars, as you would expect from self-enrichment. 
To show where the MSTO and subgiant branch stars lie, all objects from 
the ``cleaned'' sample within the $[Fe/H]_{calc}$ range with no restriction
on colors or uncertainties are shown as tiny circles.

{\bf Figure 10:}
Color-magnitude diagrams of the main-sequence turn-off region of
\wcen , shown with isochrones corresponding to the mean $[Fe/H]_{calc}$
of that group. All stars shown as filled triangles, open squares and open circles
 have $\sigma_V = \sigma_{b-y} \leq 0.01$, and $ \sigma_{m_1} \leq 0.02$.
\bf a: \rm filled triangles have $-1.2<[Fe/H]<-0.5$, isochrones correspond to
$[Fe/H]=-1.14$, mean age $<8$~Gyr;
\bf b: \rm open squares have $-1.6<[Fe/H]<-1.2$, isochrones correspond to
$[Fe/H]=-1.41$, mean age $9.4\pm 1.4$~Gyr;
 \bf c:\rm
open circles have $-2.2<[Fe/H]<-1.6$,  isochrones correspond to
$[Fe/H]=-1.84$, mean age $12.7\pm 1.5$~Gyr.
To show where the MSTO and subgiant branch stars lie, all objects from
the ``cleaned'' sample within the $[Fe/H]_{calc}$ range with no restriction
on colors or uncertainties are shown as tiny circles.

{\bf Figure 11:}
Color-magnitude diagrams of the main-sequence turn-off region of
NGC~6397, shown with isochrones corresponding to the mean $[Fe/H]_{calc}$
of the custer, $[Fe/H]=-1.84$. All stars shown as filled triangles, open squares and open circles
 have $\sigma_V = \sigma_{b-y} \leq 0.012$, and $ \sigma_{m_1} \leq 0.023$.
\bf a: \rm filled triangles have $-1.2<[Fe/H]<-0.5$, isochrones correspond to
$[Fe/H]=-1.84$, there are few objects within the calbration range with very good photometry here, as we
expected;
\bf b: \rm open squares have $-1.6<[Fe/H]<-1.2$, isochrones correspond to
$[Fe/H]=-1.84$, consistent with a mean age of around $11.7\pm 0.9$~Gyr;
 \bf c:\rm
open circles have $-2.2<[Fe/H]<-1.6$,  isochrones correspond to
$[Fe/H]=-1.84$, mean age $11.8\pm 1.4$~Gyr.
To show where the MSTO and subgiant branch stars lie, all objects from
the ``cleaned'' sample within the $[Fe/H]_{calc}$ range with no restriction
on colors or uncertainties are shown as tiny circles.

{\bf Figure 12:} Age spreads and metallicity spreads in the two clusters.
\bf a: \rm Plot of mean age against calculated [Fe/H] for \wcen . We divided
the samples  into groups with calculated $[Fe/H] = -1.0 \pm
0.1; -1.15 \pm 0.1; -1.3 \pm 0.1; -1.4 \pm 0.1; -1.55 \pm 0.1; -1.6 \pm 0.1;
-1.7 \pm 0.1; -1.82 \pm 0.1; -2.0 \pm 0.1; -2.15 \pm 0.1$. We plotted these
groups (with a narrow metallicity range) 
on a CMD. We estimated
the age of each star individually from the VandenBerg \etal\ (1999) grid of
isochrones corresponding to the mean metallicity of
that group, between $3 < M_V < 4$~mag. (where the isochrones are almost vertical). 
\it Here, we see that the age spread in \wcen\ is 2--4~Gyr. \rm\
Uncertainties in [Fe/H] are estimated from the Malyuto calibration to be $\sim \pm 0.15$, and
the uncertainty in the age is the standard deviation of the mean age within each metallicity group.
\bf b: \rm  Plot of mean age against calculated [Fe/H] for NGC~6397.
We divided
the samples  into groups with calculated $[Fe/H] = -1.0 \pm
0.1; -1.15 \pm 0.1; -1.3 \pm 0.1; -1.4 \pm 0.1; -1.55 \pm 0.1; -1.6 \pm 0.1;
-1.7 \pm 0.1; -1.82 \pm 0.1; -2.0 \pm 0.1; -2.15 \pm 0.1$. We plotted these
groups (with a narrow metallicity range)
on a CMD. We estimated
the age of each star individually from the VandenBerg \etal\ (1999) grid of
isochrones for $[Fe/H]=-1.84$, between $3 < M_V < 4$~mag. (where the isochrones
are almost vertical). NGC~6397 stars can only be fit by the average-metallicity
isochrones.
We see that there is no strong evidence that the metal-poor stars are older
than the metal-rich stars, and the  
sample of stars here have a weighted mean age of $11.7 \pm 0.1$~Gyr.
The unweighted mean for the whole
sample is $12.0 \pm 0.7$~Gyr.

\newpage
\centerline{\Large Table Captions}

\bf Table 1: \rm CTIO 0.9-m CCD Frames.

\bf Table 2: \rm Sample of Objects in $\omega$ Cen.

\bf Table 3: \rm Objects in the Off-Cluster Field.

\bf Table 4: \rm Sample of Objects in NGC 6397.

\newpage
\begin{references}
\parindent 0em

\ref Alcaino, G., Liller, W., Alvarado, F., Kravtsov, V., Ipatov, A., Samus, N., $\&$ Smirnov, O. 1997 \aj\ 114, 1067.

\ref Alcaino, G.,  $\&$ Liller, W. 1987, \aj\ 94, 1585

\ref Anthony-Twarog, B.J., $\&$ Twarog, B.A. 1994, \aj\ 107, 1577.

\ref Anthony-Twarog, B.J., Twarog, B.A., $\&$ Craig, J. 1993, \pasp\ 107, 32.

\ref Anthony-Twarog, B.J., Twarog, B.A.,  $\&$ Suntzeff, N.B. 1992, \aj\, 103, 1264.

\ref Astronomer Royal, Royal Observatory Annals, \it Number 2, \rm ``Studies of the Globular
Cluster \wcen\ I. Catalogue of Magnitudes and Proper Motions,'' 1966, 
London: Her Majesty's Stationary
Office.

\ref Bates, B., Wood, K.D., Catney, M.G., $\&$ Gilheany, S. 1992, \mnras\ 254, 221.

\ref Bergbusch, P.A., $\&$ VandenBerg, D.A. 1992, \apjs\ 81, 163.

\ref Bergbusch, P.A., $\&$ VandenBerg, D.A. 1997, \aj\ 114, 2604.

\ref Briley, M.M., Hesser, J.E., $\&$ Bell, R.A. 1991, \apj\ 373, 482.

\ref Brown, J.H., Burkert, A., Truran, J.W., 1991, \apj\ 376, 115.
249, L13.

\ref Brown, J.A., Wallerstein, G., $\&$ Oke, J.B. 1990 \aj\ 100, 1561.

\ref Butler, D., Dickens, R.J., $\&$ Epps, E. 1978, \apj\ 225, 148.

\ref Cannon, R.D., Croke, B.F.W., Bell, R.A., Hesser, J.E., $\&$ Stathakis,
R.A. 1998, \mnras\ 298, 601.

\ref Crawford, D.L., and Mandwewala, N. 1976, \pasp\ 88, 917.

\ref De Marchi, G. 1999, \aj\ 117, 303.

\ref Dickens, R.J., $\&$ Wooley, R. 1967, Royal Observatory bulletins.
Series E; no.128, London:H.M.S.O., p.255.

\ref Dinescu, D.I., Girard, T.M., van Altena, W.F. 1999, \aj\ 117, 1792.

\ref Freeman, K.C. 1993, in A.S.P. Conf. Ser. Vol. 48, \it ``The Globular
Cluster-Galaxy Connection,'' \rm Edited by G.H.Smith $\&$ J.P.Brodie, p.27.

\ref Folgheraiter, E.L., Penny, A.J., Griffiths, W.K., $\&$ Dickens, R.J.
1995, \mnras\ 274, 407.

\ref Gehrels, N. 1986, \apj\ 303, 336.

\ref Gratton, R.G., Fusi Pecci, F., Carretta, E., Clementini, G., Corsi, C.E.,
$\&$ Lattanzi, M. 1997, \apj\ 491, 749.

\ref Grebel, E. K., $\&$  Richtler, T. 1992  \aap\ 253, 359.

\ref Hilker, M., $\&$ Richtler, T. 1999, \aap ,  in \it Proceedings of the
35th Liege International Astrophysics Colloqium: ``The Galactic Halo: from
Globular Clusters to Field Stars, in press (astro-ph/9910370).

\ref Kilkenny, D., $\&$ Laing, J.D. 1992, \mnras\ 225, 308.

\ref Kurucz, R.L. 1992, \rm Kurucz  CD-ROM 19, Solar Abundance Model Atmospheres
(Cambridge: SAO).

\ref Evans, T.L. 1983, \mnras\ 204, 975.

\ref Majewski, S.R., Patterson, R.J., Dinescu, D.I., Johnson, W.Y., 
Ostheimer, J.C., Kunkel, W.E., Palma, C. 1999, in \it Proceedings of the
35th Liege International Astrophysics Colloqium: ``The Galactic Halo: from
Globular Clusters to Field Stars, in press (astro-ph/9910278).

\ref Malyuto, V. 1994, \aaps\ 108, 441.

\ref Mighell, K.J., Sarajedini, A., $\&$ French, R.S. 1998, \aj\ 116, 2395.

\ref Morgan, S. $\&$ Lake, G. 1989, \apj\ 339, 171.

\ref Mukherjee, K., Anthony-Twarog, B.J.,  $\&$ Twarog, B.A.  1992, PASP 104, 561.

\ref Nissen, P.E. 1981, \aap\ 97, 145.

\ref Norris, J.E., $\&$ Da Costa, G.S. 1995, \apj\ 447 680.

\ref Norris, J.E., Freeman, K.C., $\&$ Mighell, K.J. 1996, \apj\ 462, 241.

\ref Norris, J.E., Freeman, K.E, Mayor, M. $\&$ Seitzer, P. 1997, \apj\ 487, 187.

\ref Padoan, P., Jimenez, R, $\&$ Jones, B. 1997 \mnras\ 185, 711.

\ref Parmentier, G., Jehin, E., Magain, P., Neuforge, C., Noels, A. $\&$
Thoul, A.A. 1999, \aap\  in press (astro-ph/9911258). 

\ref Reid, N.I. 1997, \aj\ 114, 161.

\ref Reid, N.I., $\&$ Gizis, J.E. 1998, \aj\ 116, 2929.

\ref Richter, P., Hilker, M., $\&$ Richtler, T. 1999, 
\aap\ in press (astro-ph/9907200).

\ref Schuster, W.J., $\&$ Nissen, P.E. 1989, \aap\ 222, 69.

\ref Schuster, W.J., Nissen, P.E., Parrao, L., Beers, T.C., $\&$ Overgaard, L.P.
1996, \aaps\ 117, 317.

\ref Stetson, P.B., Davis, L.E., $\&$ Crabtree, D.R. 1990, in \it CCDs in Astronomy,
\rm
ASP Conf. Ser. 8, 289, Ed. G.H. Jacoby.

\ref \strom , B. 1966, ARAA, 4, 433.

\ref Suntzeff, N.B. $\&$ Kraft, R.P. 1996, \aj\ 111, 1913.

\ref VandenBerg, D.A., $\&$ Bell, R.A. 1985, \apjs\ 58, 561.

\ref VandenBerg, D.A., Swenson, F.J., Rogers, F.J., Iglesias, C.A., $\&$ Alexander,
D.R. 1999, in preparation.

\ref Van den Bergh, S. 1993, \apj\ 411, 178.

\ref Vanture, A.D., Wallerstein, G., $\&$ Brown, J.A. 1994, PASP 106, 835.

\ref Wood, K.D., $\&$ Bates, B. 1994, \mnras\ 267, 660.

\ref Wood, K.D., $\&$ Bates, B. 1993, \apj\ 417, 572.

\end{references}
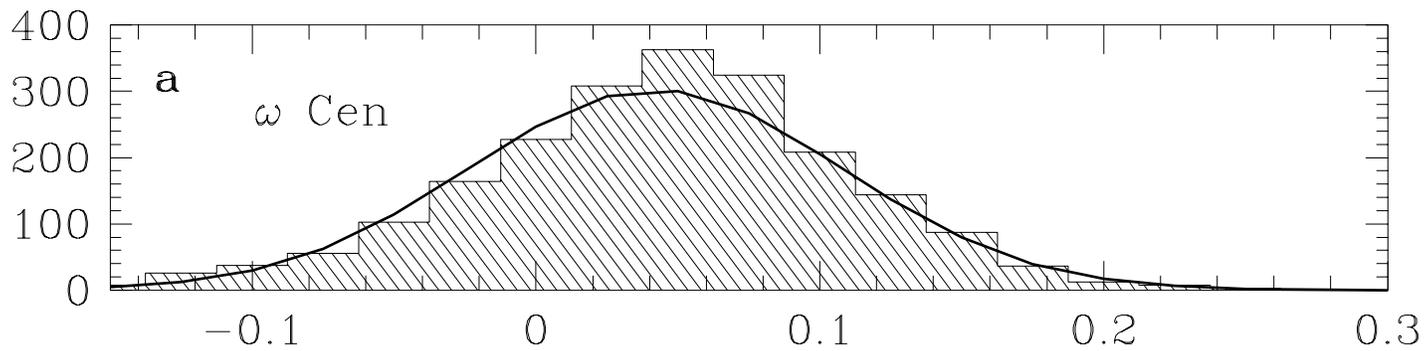
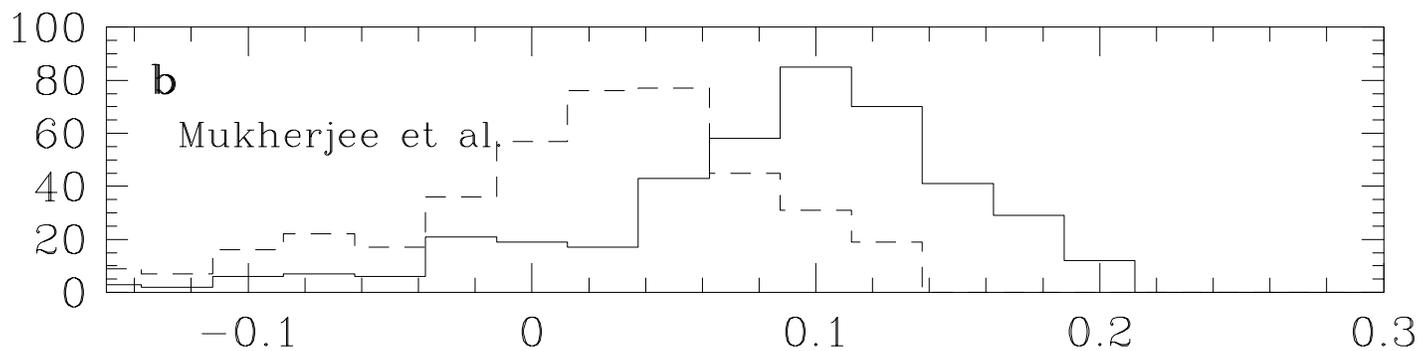
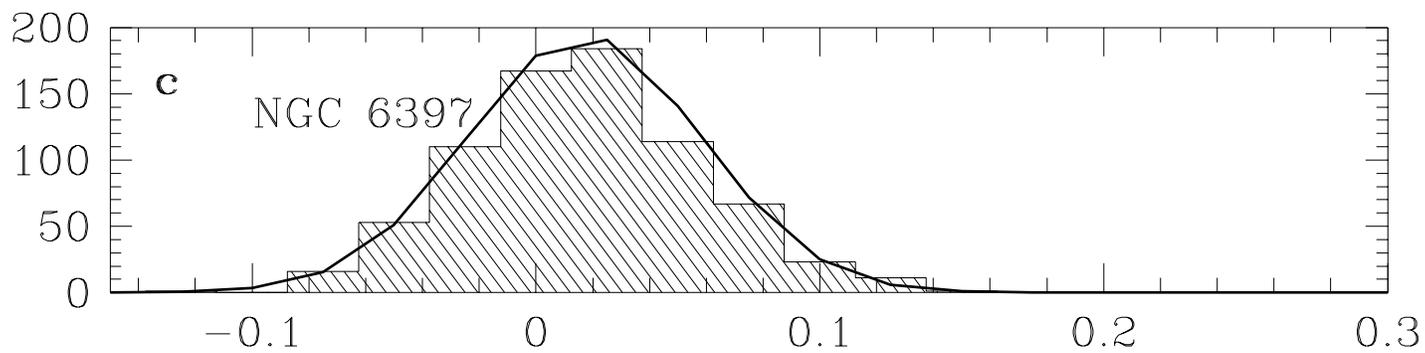
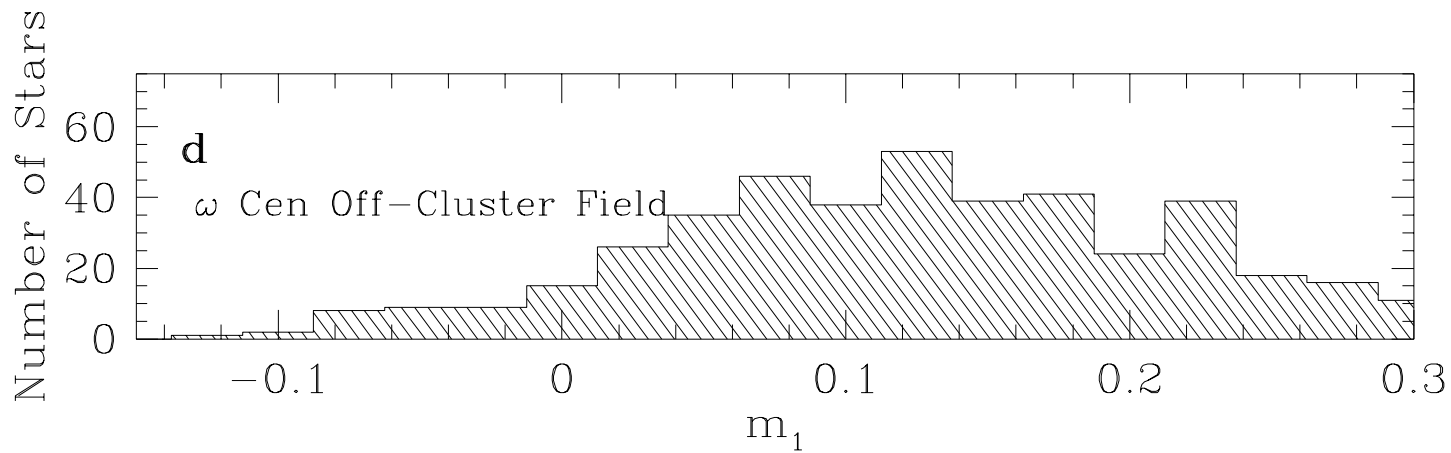
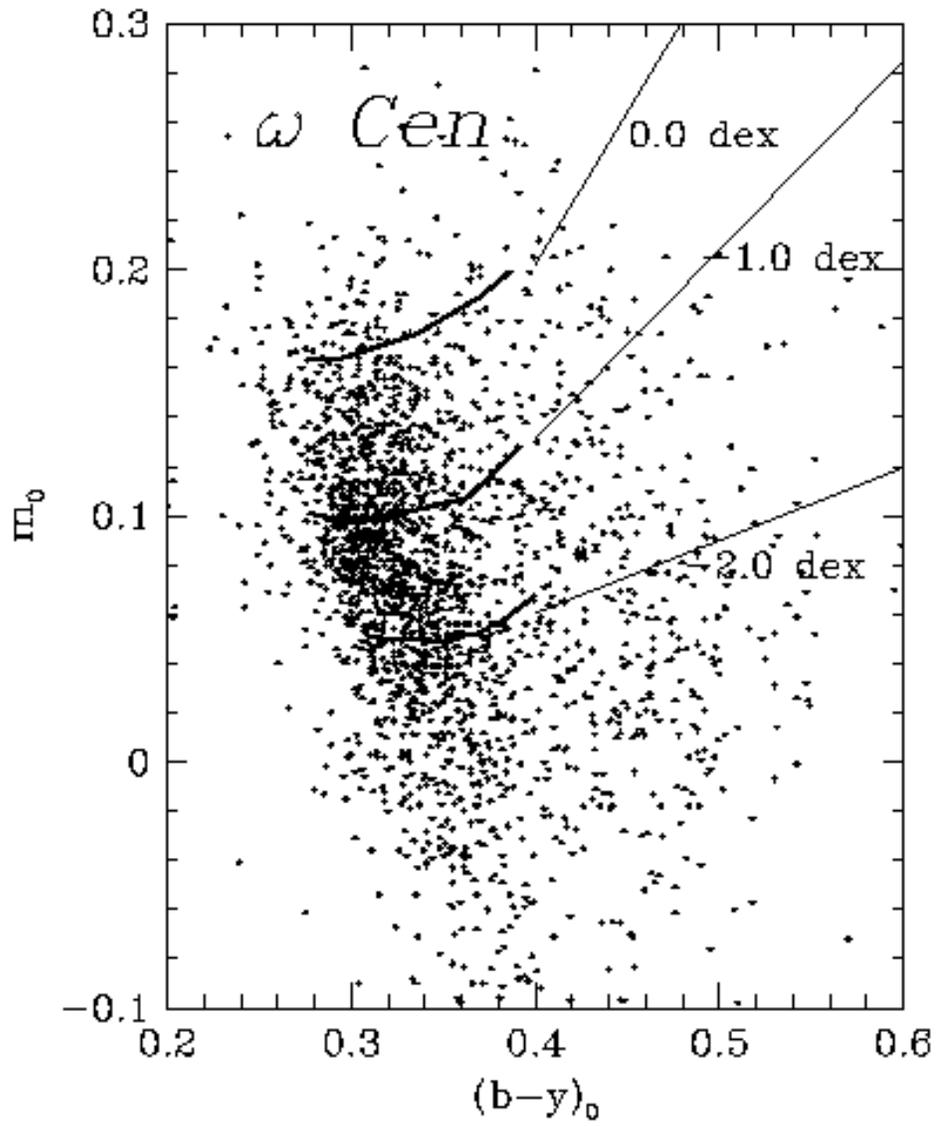
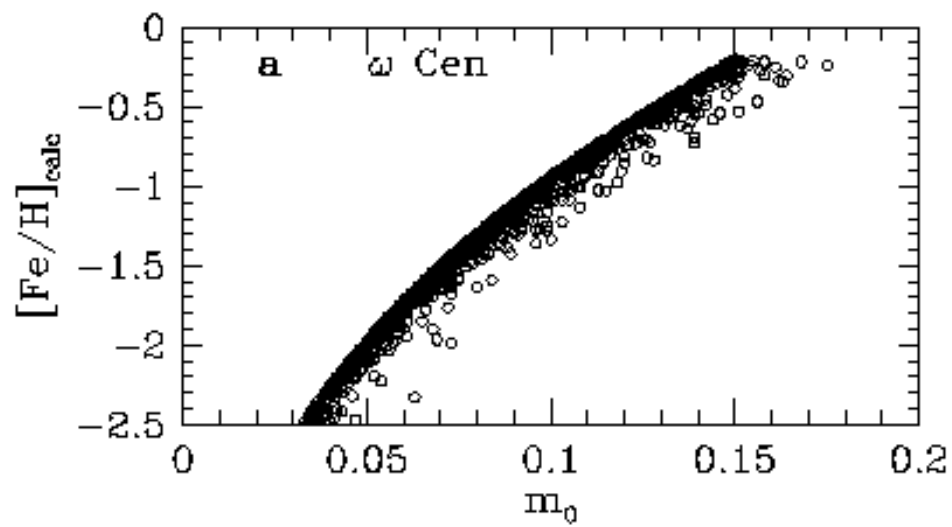
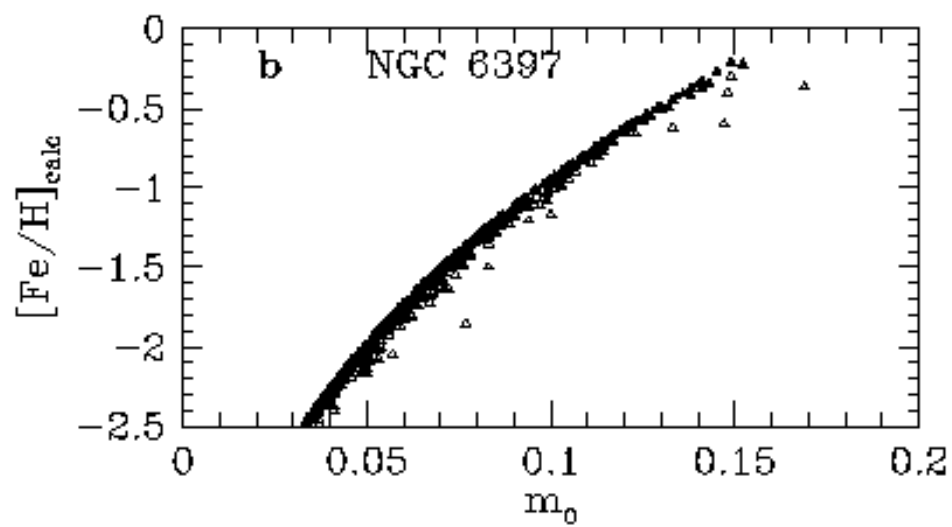
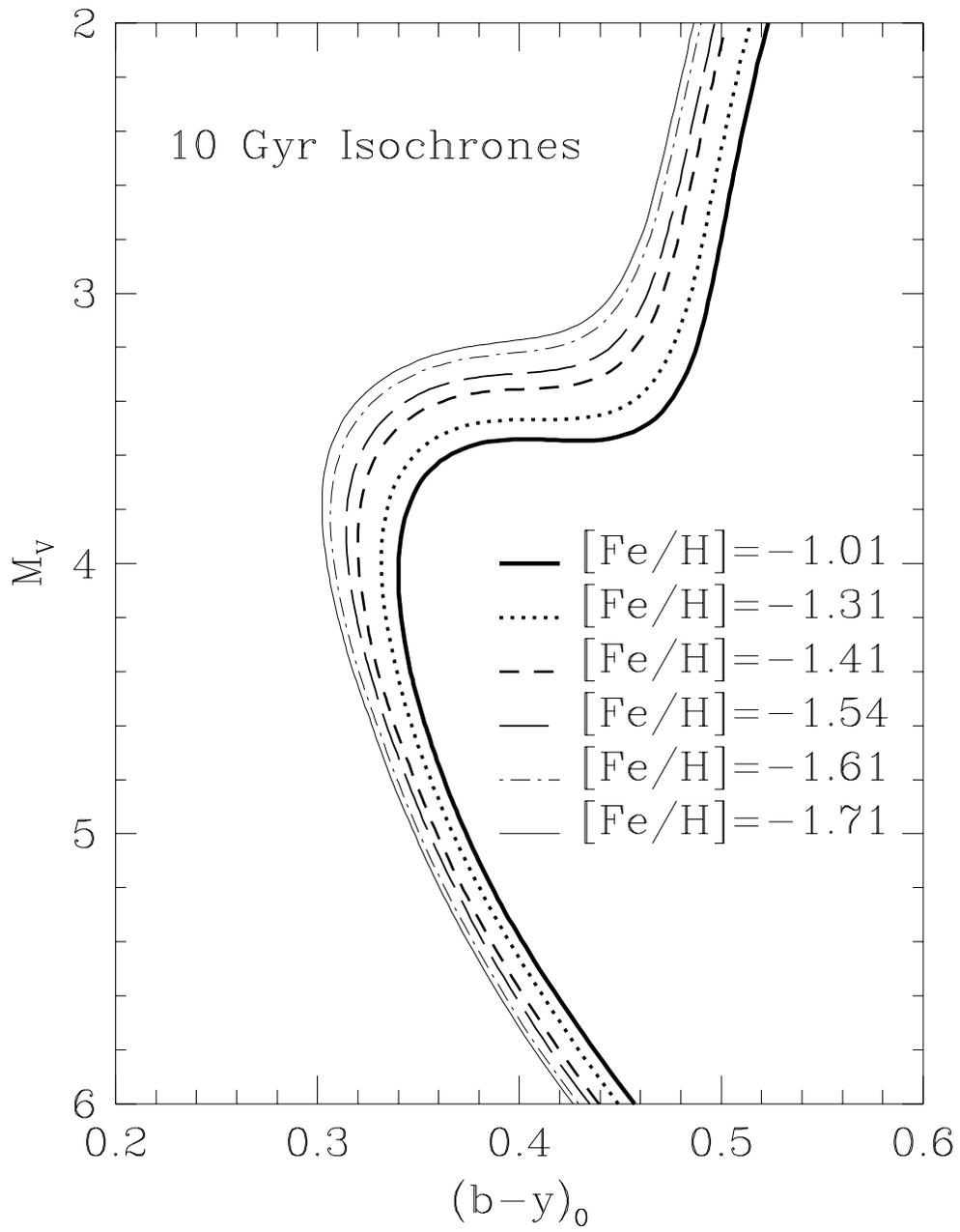
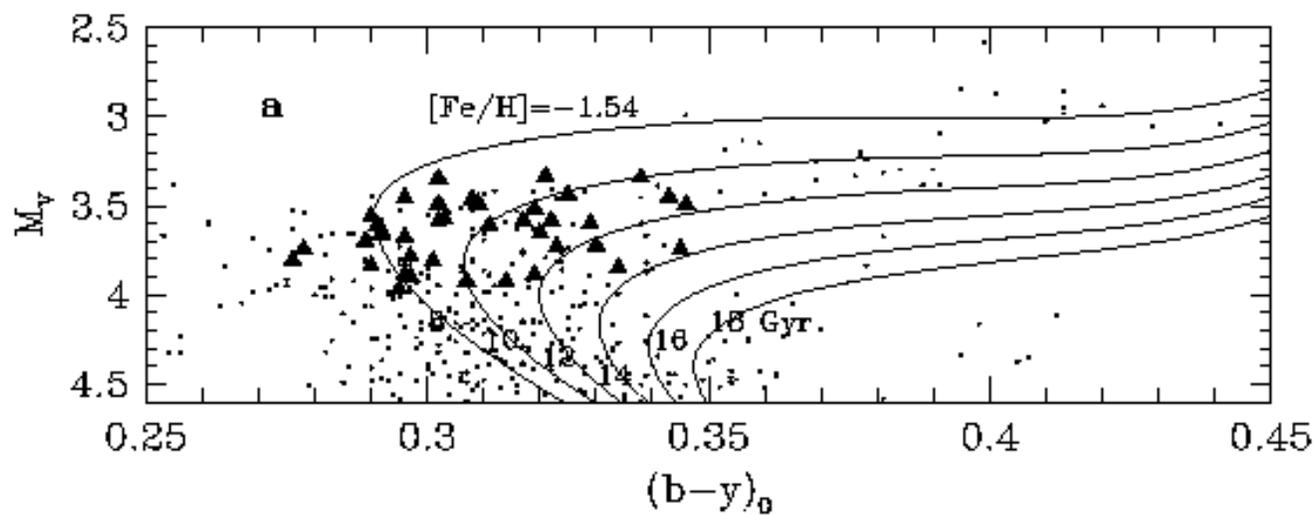
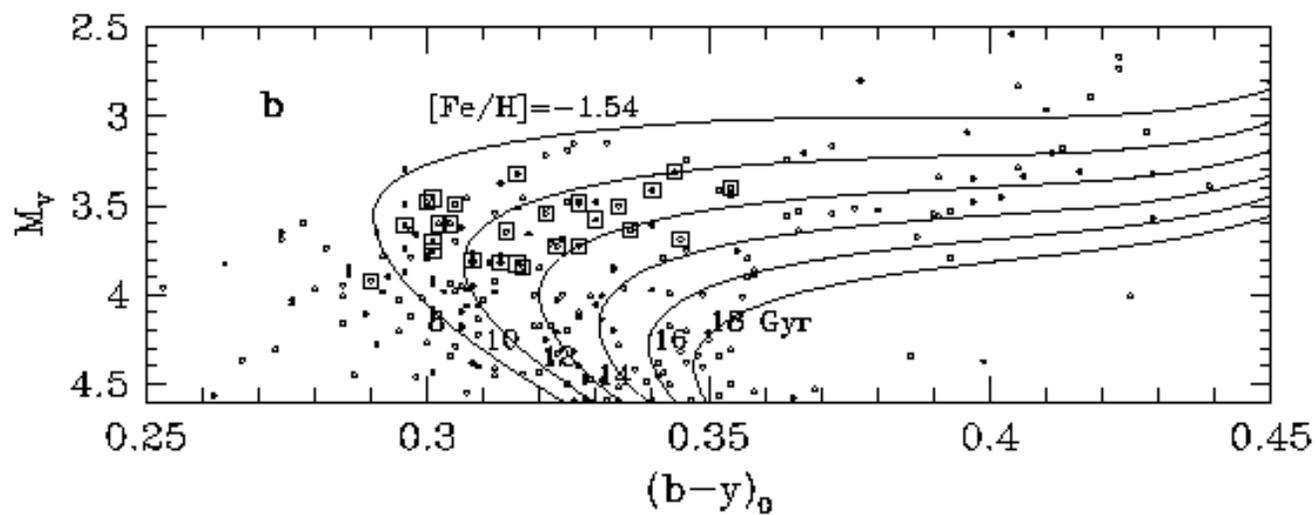
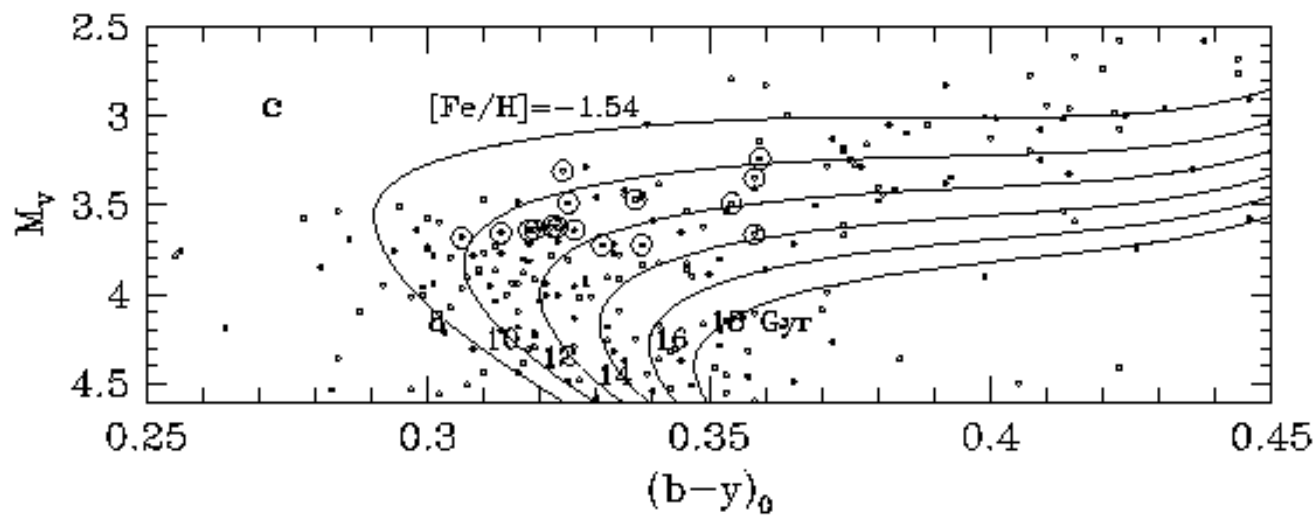
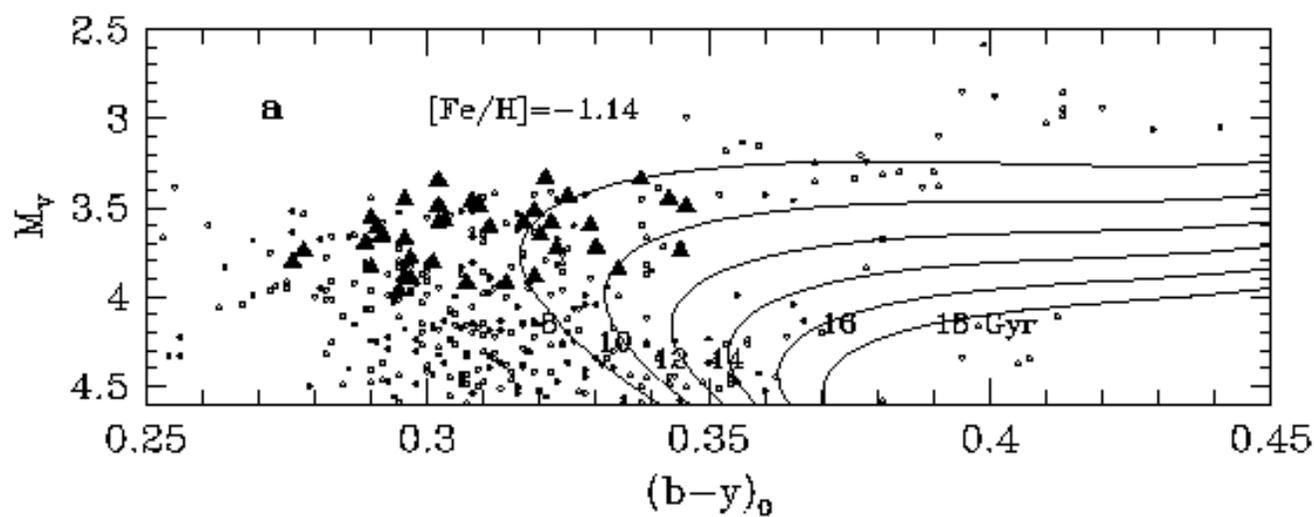
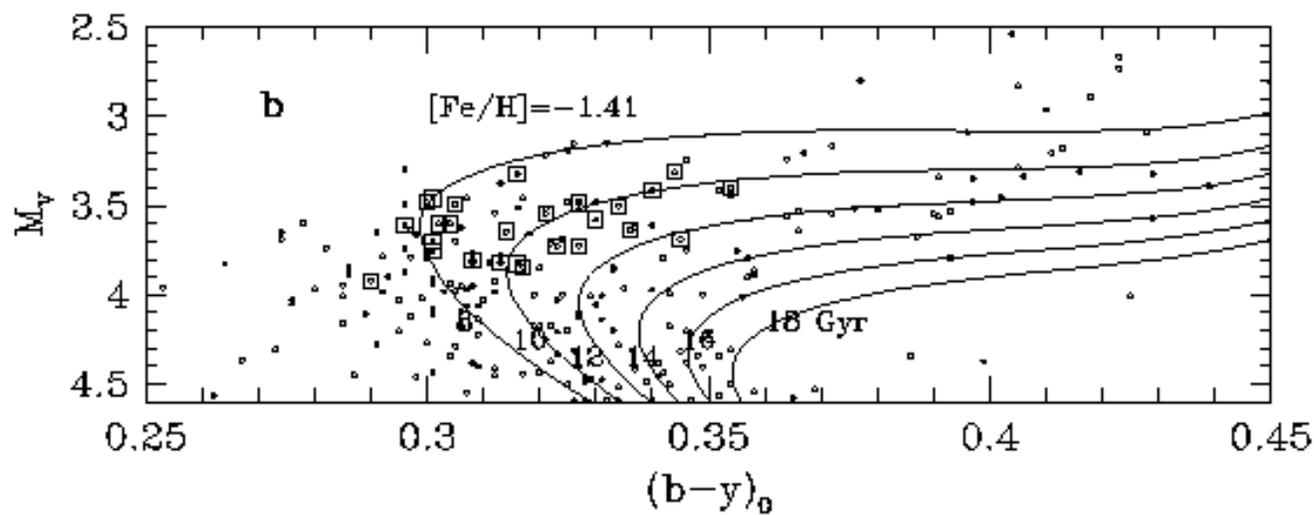
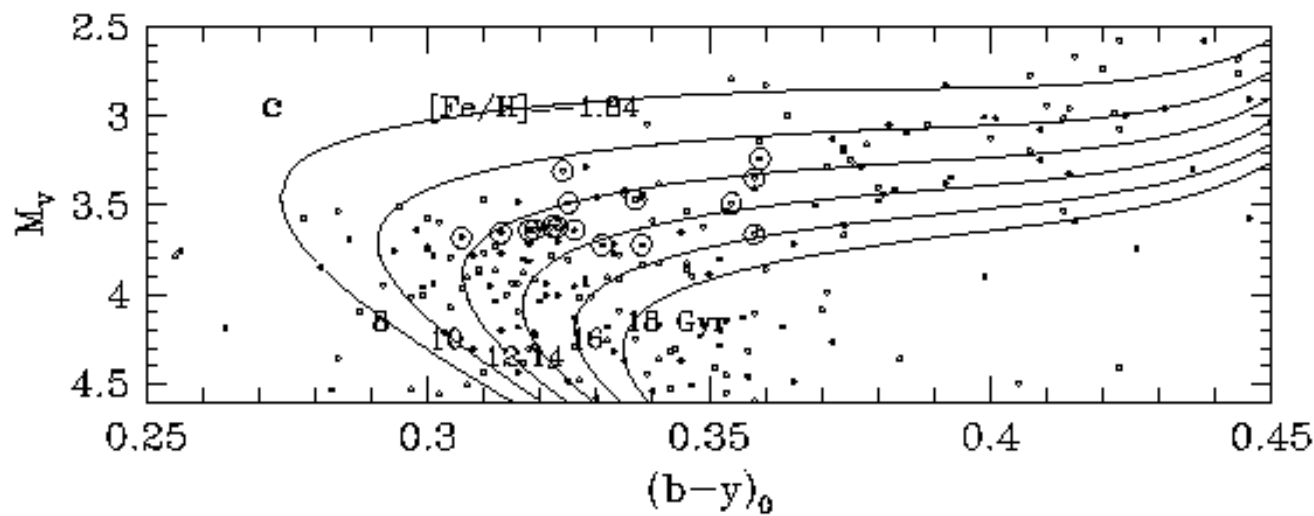
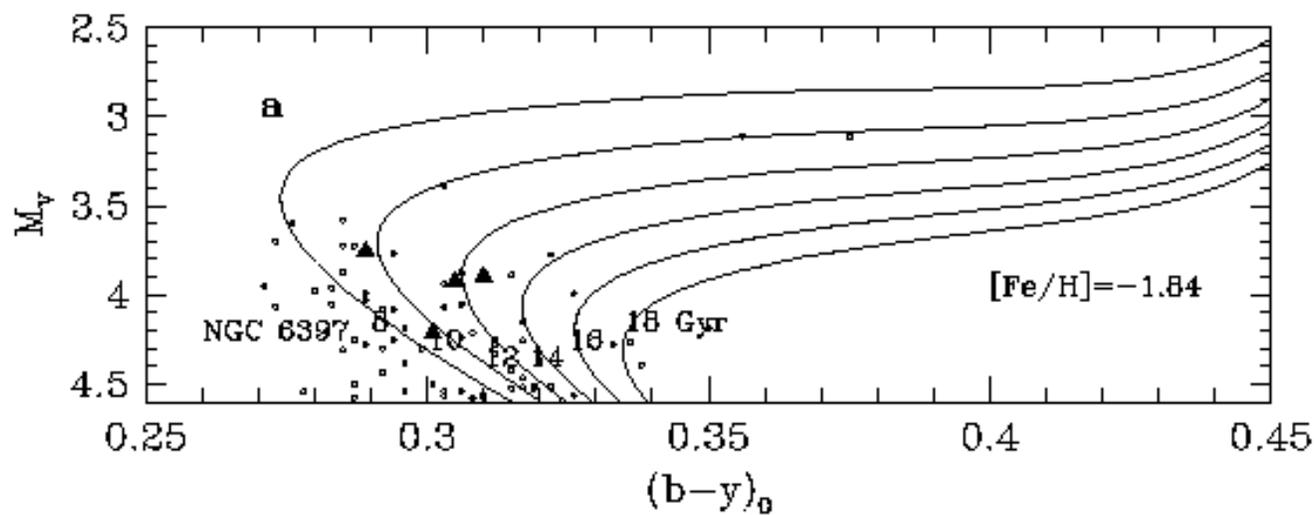
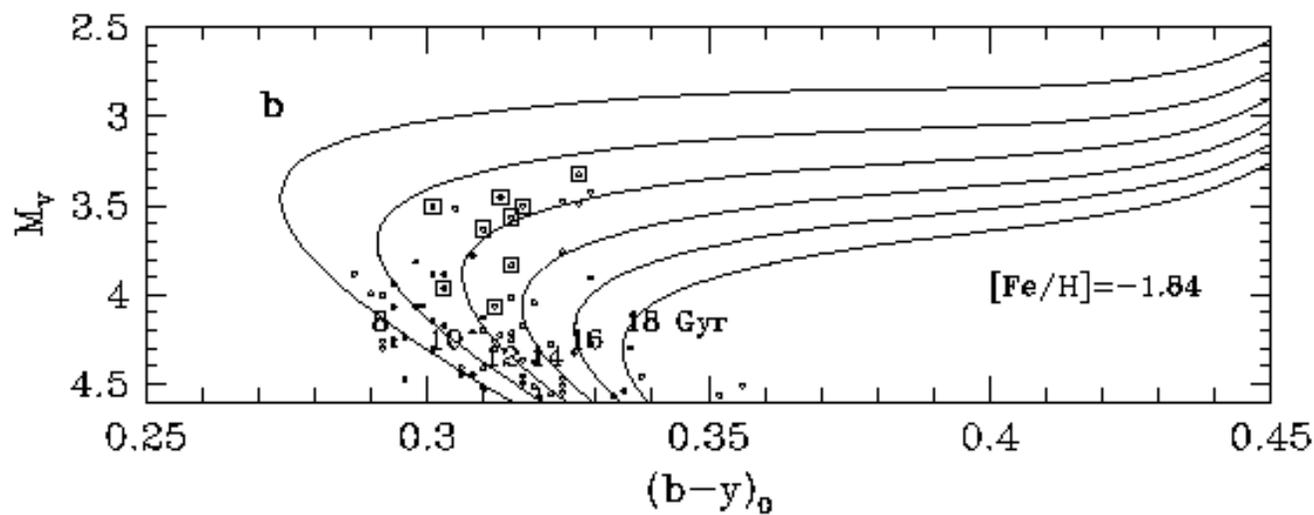
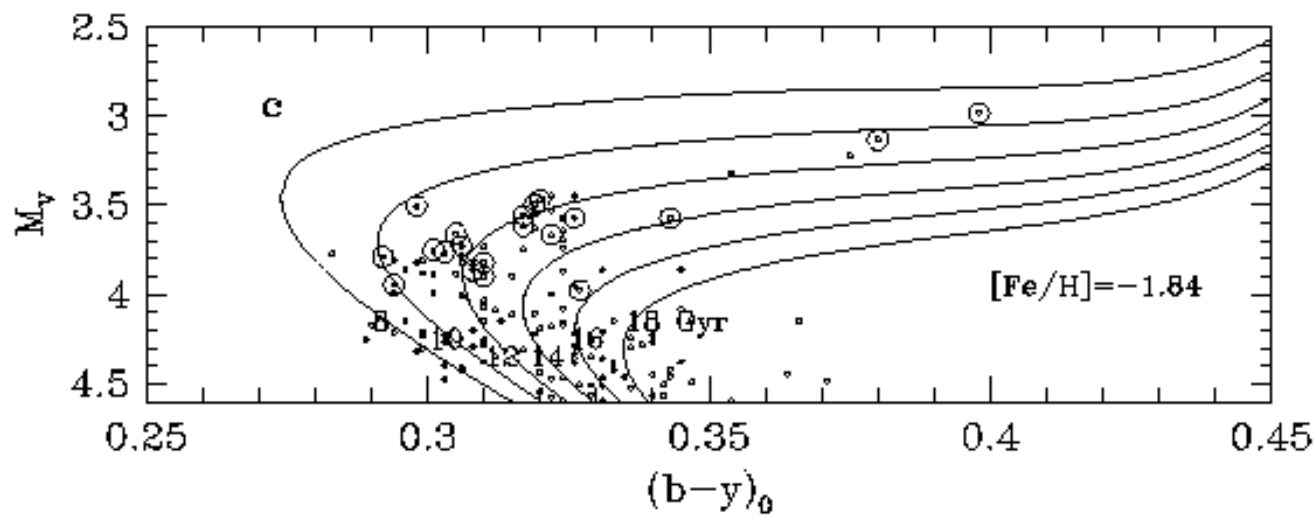
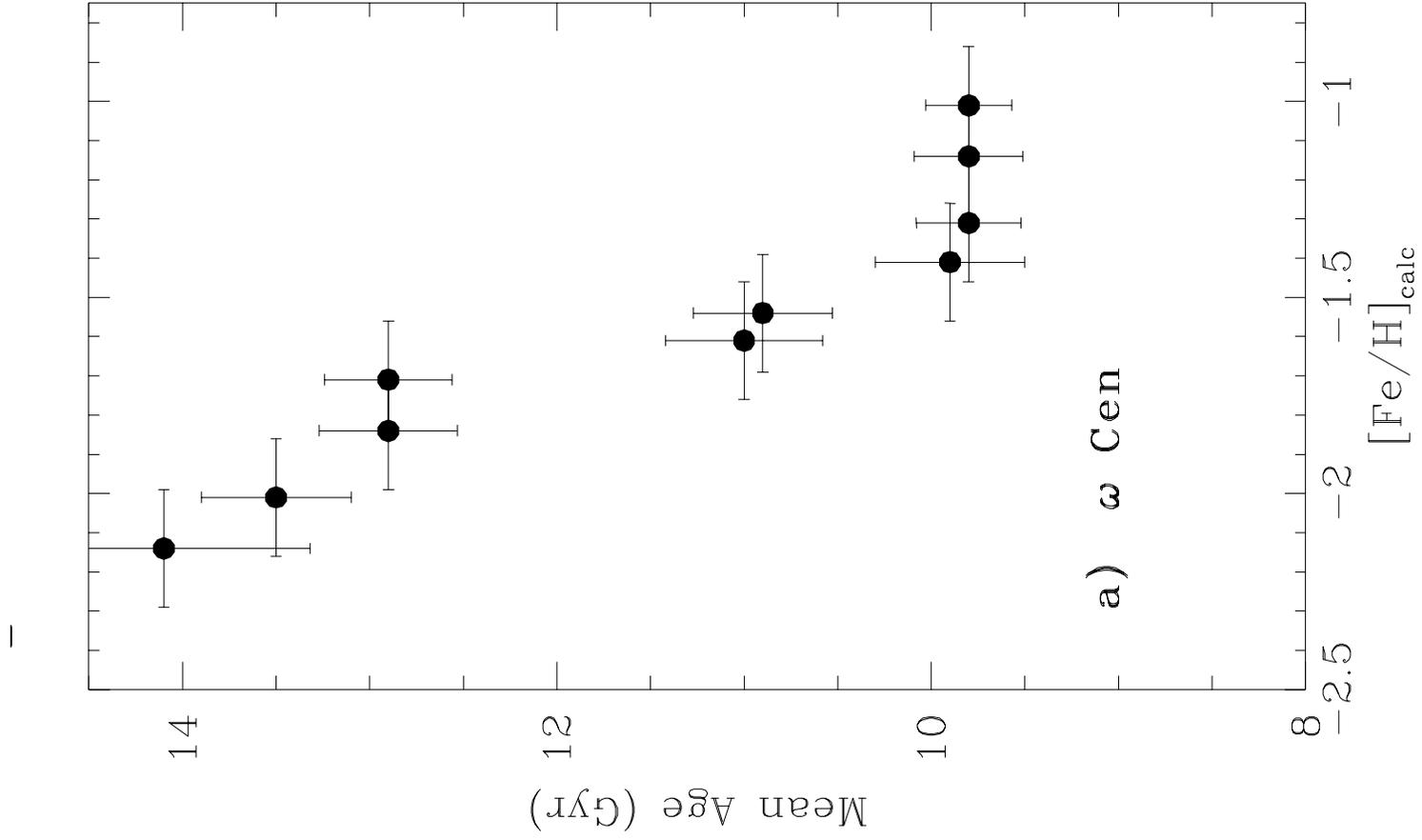
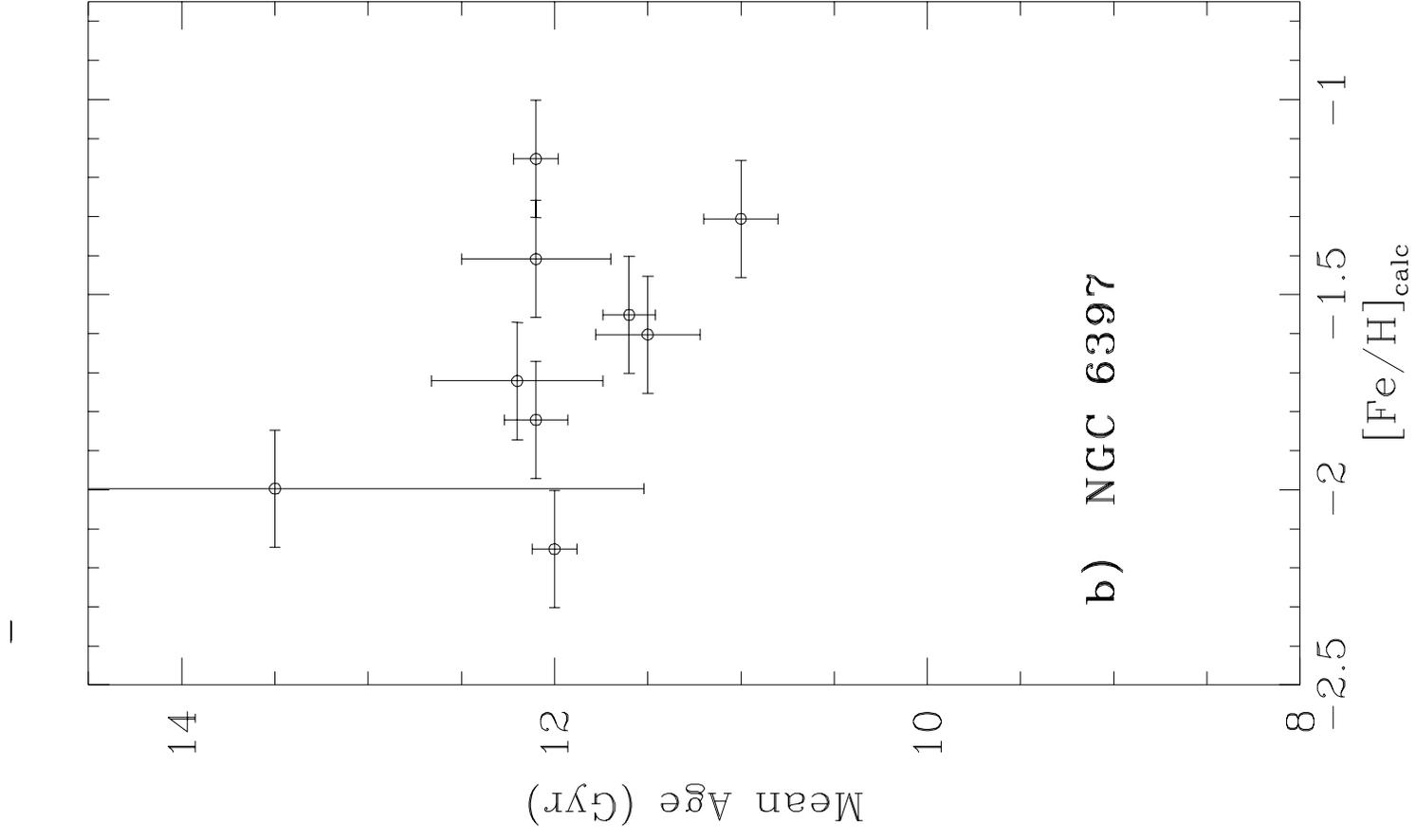
\end{document}